# Two Shifts for Crop Mapping: Leveraging Aggregate Crop Statistics to Improve Satellite-based Maps in New Regions

Dan M. Kluger[a,*], Sherrie Wang[b,c], David B. Lobell[b]

[a]  Department of Statistics, Sequoia Hall, Mail Code 4095, 390 Jane Stanford Way, Stanford
     University, Stanford, CA 94305-4020, United States of America

[b] Department of Earth System Science and Center on Food Security and the Environment, Encina
    Hall East, 616 Jane Stanford Way, Stanford University, Stanford, CA 94305-4205, United
    States of America

[c] Institute for Computational and Mathematical Engineering, Huang Engineering Building, 475
    Via Ortega, Suite B060, Stanford, CA 94305-4042, United States of America

[*]  Corresponding author

E-mail addresses: kluger@stanford.edu (Dan M. Kluger), sherwang@stanford.edu (Sherrie
Wang), dlobell@stanford.edu (David B. Lobell)










Abstract: Crop type mapping at the field level is critical for a variety of applications in agricultural monitoring, and satellite imagery is becoming an increasingly abundant and useful raw input from which to create crop type maps. Still, in many regions crop type mapping with satellite data remains constrained by a scarcity of field-level crop labels for training supervised classification models. When training data is not available in one region, classifiers trained in similar regions can be transferred, but shifts in the distribution of crop types as well as transformations of the features between regions leads to reduced classification accuracy. We present a methodology that uses aggregate-level crop statistics to correct the classifier by accounting for these two types of shifts. To adjust for shifts in the crop type composition we present a scheme for properly reweighting the posterior probabilities of each class that are output by the classifier. To adjust for shifts in features we propose a method to estimate and remove linear shifts in the mean feature vector. We demonstrate that this methodology leads to substantial improvements in overall classification accuracy when using Linear Discriminant Analysis (LDA) to map crop types in Occitanie, France and in Western Province, Kenya. When using LDA as our base classifier, we found that in France our methodology led to percent reductions in misclassifications ranging from 2.8% to 42.2% (mean=21.9%) over eleven different training departments, and in Kenya the percent reductions in misclassification were 6.6%, 28.4%, and 42.7% for three training regions. While our methodology was statistically motivated by the LDA classifier, it can be applied to any type of classifier. As an example, we demonstrate its successful application to improve a Random Forest classifier.






## 1) Introduction

Crop type mapping at the field resolution is a prerequisite to mapping farm management and yield outcomes at large spatial scale. This task is all the more urgent in a time when populations in food-insecure regions continue to increase and climate change is predicted to adversely affect global agriculture.

Traditionally, crop type information has been obtained from field surveys and censuses, but such surveys are expensive and time consuming to conduct. Therefore, in many countries, the surveys are often constrained to small geographical regions. With the increasing abundance and availability of satellite data, there is an opportunity to augment these surveys by creating low-cost crop type maps using remotely sense spectral band time series data (Waldner *et al.*, 2015). Efforts to augment data from limited field surveys has been aided by the recent success of various machine learning methods including Neural Networks (Cai *et al.*, 2018; Fu *et al.,* 2017).

There are still major challenges, however, in using machine learning methods to create crop type maps. First, these methods have seen limited success in settings where there is not abundant labeled data, as is the case in many countries (Defourny *et al.,* 2019). Second, since the labeled data is often constrained to a small geographical region, using machine learning methods to extrapolate crop type maps in regions outside of the geographical training region can be plagued by heterogeneity of agricultural settings. While transferring machine learning classifiers to neighboring regions can help, this approach has seen limited success so far. In Wang *et al.* (2019), the test set transfer accuracy of Random Forests trained and tested on distinct states in the United States was high when the two states had similar growing degree days but performed





poorly when the growing degree days differed by a lot. Further, Random Forest transfer accuracy tends to be lower when the proportions of each crop type on the training region are very different than those on the test region (Wang *et al*., 2019, Waldner *et al*., 2017).

Due to the evident need to account for regional discrepancies when transferring machine learning prediction tools between regions, a few deep learning-based approaches have been proposed to reduce accuracy loss when transferring a classifier trained on remote sensing data to new geographical regions. Approaches that involve removing the last layer of a pretrained Neural Network and building a Neural Network on top of the removed layer have seen success in remote sensing applications (Wang *et al.*, 2018, Pires de Lima and Marfurt, 2019); however, these approaches require some training labels from the target dataset or region. Another type of deep learning-based approach is adversarial domain adaptation, which has shown recent success in remote sensing tasks of vehicle detection (Koga *et al.,* 2020), urban land cover mapping (Fang *et al.*, 2019), and even satellite-based crop classification (Elshamli *et al.*, 2017). Despite their promise and success, these methods are computationally intensive and none of them leverage prior information about the distribution of class labels in the target region.

The goal of this paper is to present and test a methodology for improving the accuracy of transfer learning in multi-class crop classification settings where only one region has labeled data but all regions have aggregate-level crop statistics. The methodology corrects for two types of shifts between the training and testing region. The first type is a shift in the distribution of crop types between regions, which has been observed to lower classification accuracy in the test region (Waldner *et al.*, 2017; Wang *et al.*, 2019). The second type is a systemic transformation of the





features between regions, often caused by differences in farm management, growing degree days, or other climatic variables. For example, growing seasons can shift, shrink, extend as one moves to new climatic regimes. In addition, the value of spectral bands at the peak of the growing season can shift up or down if the overall suitability of conditions differs between regions. We call the methods presented in this paper for correcting for each type of shift "Prior Shift Adjustment" (PSA) and "Feature Shift Adjustment" (FSA) respectively. The two methods are easily combinable to correct a classifier for both types of shifts simultaneously.

We emphasize that the methodology here relies on using aggregate-level crop type statistics to estimate the distribution of the crop labels in each region. Aggregate-level crop statistics are often available even in settings where field-level crop type data are unavailable. For example, You *et al.,*2014 incorporated large amounts of subnational aggregate-level data from Agro-MAPS and other sources into their Spatial Production Allocation Model, which at the time of this writing, uses aggregate-level crop statistics from at least 60% of level 1 administrative units and from at least 35% of level 2 administrative units globally (MapSPAM methodology, n.d.).

Because of the availability of aggregate-level crop statistics, crop type mapping is a machine learning-based classification setting where it is commonly possible to estimate the distribution of the true classes on an unlabeled target dataset. In machine learning, having prior knowledge of the distribution of the classes on an unlabeled target dataset is a relatively uncommon luxury; therefore, methods that exploit this luxury are either not widely taught or are underdeveloped. To our knowledge, no feature shift adjustment method has been proposed that leverages prior information on the distribution of the classes in the target unlabeled dataset. Thus, it is important





to explore and develop new methods that leverage prior information about the distribution of the classes within the target unlabeled dataset, as exploring such methods can lead to improved crop type maps.

To make progress toward this goal, this paper focuses on experiments in two contrasting settings: Occitanie, France and Western Province, Kenya. In each case, we test the performance of models trained in one part of the region and then applied in the remaining parts of the region. The labeled dataset in Occitanie, France spans 13 different departments and is both large and accurate, making it an ideal dataset to verify the validity and accuracy of the methodology we present. The dataset from Western Province, Kenya is more representative of a setting where field-level crop labels are scarce and constrained to small geographical regions, and therefore represents a setting where such a methodology would likely be used.

The structure of this paper is as follows. In Section 2 we present the prior shift and feature shift adjustment methods. The first is a well-established method to correct a classifier for differences in the distributions of classes between the training and test regions (Saerens *et al.*, 2002). The second method corrects for a shift in the mean value of each feature as one moves from the training region to the test region, without removing the component of feature mean shift that is driven by differing crop type distributions between the two regions. In Section 3, we describe our two datasets in detail and the data processing steps that were taken. In Section 4, we present the accuracy of our methods when training on one region and testing on other regions for two types of classifiers, LDA and Random Forest. We also compare our results with those of transfer learning approaches that use data balancing or standardization techniques. In Section 5, we





interpret our results and discuss why our methodology is most suitable for LDA classifiers but still helpful for other machine learning classifiers. We also discuss the benefits of picking a diverse region to train on, if the option is available.

## 2) Prior Shift and Feature Shift Adjustment methodology

### 2.1) Prior Shift Adjustment

In machine learning-based classification tasks, the distribution of test set labels can be very different from that of the training set, leading to suboptimal classification accuracy. To pinpoint the reason for this reduced classification accuracy, we briefly describe the underlying steps often involved in machine learning classification. Many machine learning classifiers (e.g. LDA, Multiple Logistic Regression, Neural Networks, etc.) use the training data features and labels to determine a function that maps the features of an observation to the estimated posterior probabilities that the observation is in each class. To estimate the label, the classifier chooses the class with the highest estimated posterior probability. This function, however, gives posterior probability estimates for points that are drawn from the training set distribution, whereas ideally the function would give posterior probability estimates for points drawn from the test set distribution. Therefore, problems can arise in supervised learning when the proportion of test set labels of each class is very different from that of the training set (see Waldner *et al.*, 2017 for a reference highlighting this issue in a binary crop classification setting). For example, if in the training set wheat is much more prevalent than barley, but in the test set barley is more prevalent than wheat, a classifier applied naively (i.e. without prior shift adjustments) will guess wheat in borderline cases on the test set when intuitively it should guess barley in borderline cases.





To address this issue in settings where the label distribution of the test set is known, we propose using the method presented in equation 4 of Saerens *et al.*, 2002. We present the method in the subsequent paragraph and provide a mathematical justification for the method in Appendix A.

Suppose there are $K$ different classes and let $X$ and $Y$ be the features and class label, respectively, of a randomly sampled point (from either the training set or the test set). Let $P_{Train}$ and $P_{Test}$ denote the joint distributions for $(X, Y)$ on the training set and test set, respectively. For each $k = 1, \dots, K$ let $\hat{P}_{Train}(Y = k \mid X = x)$ be the estimate of the posterior probability that a point with features $x$ is in class $k$. Such posterior probabilities are automatically returned by standard machine learning classification methods such as LDA, Multiple Logistic Regression, Neural Networks, and even Random Forest (though those for Random Forest and Neural Networks can be poorly calibrated). Further, let $\hat{P}_{Test}(Y = k)$ be an estimate of the proportion of crops in class $k$ on the test set, which is calculated using aggregate-level crop statistics (see Section 3.2), and let $\hat{P}_{Train}(Y = k)$ be the proportion of labeled training points that are in class $k$. The Prior Shift Adjusted (PSA for short) classifier presented in Equation 4 of Saerens *et al.*, 2002, is given by

$$\hat{l}_{PSA}(x) = \underset{k \in \{1, 2, \dots, K\}}{\mathrm{argmax}} \left\{ \hat{P}_{Train}(Y = k \mid X = x) \frac{\hat{P}_{Test}(Y = k)}{\hat{P}_{Train}(Y = k)} \right\}$$

According to the above formula, the PSA classifier reweights the posterior probabilities for each class by the ratio of that class's prevalence on the test set over its prevalence on the training set





and picks the class with the highest such reweighted posterior probability to be the fitted label.

Note that when the distribution of the training labels and the test labels are estimated to be the

same, the PSA classifier will be exactly the same as the unadjusted classifier which assigns a

class label of $\underset{k \in \{1,2,...,K\}}{\mathrm{argmax}} \, \hat{P}_{Train}(Y = k \mid X = x)$ to an unlabeled instance with features $x$. The above

formula for the PSA classifier involves a similar reweighting approach to Importance Weighted

Cross Validation (Sugiyama *et al.*, 2007) but has a couple of notable differences: in PSA the

weights are ratios of class probabilities, rather than ratios of feature densities, and PSA is a

method that is applied after training a base classifier, rather than a scheme for reweighting the

loss function before training the classifier.

In Appendix A, we show that the PSA method maximizes the overall accuracy when we assume

that the conditional distributions of the features given the class label are the same on the training

and test set and that the classifier properly learns such conditional distributions. Further, we

argue in Appendix A that the method is nearly optimal when these assumptions are weakly

violated.

## 2.2) Feature Shift Adjustment

When applying a classifier to a separate region than it was trained on, another major cause of

accuracy loss is that there can be a systemic change in feature space between regions. This

transfer learning hurdle is sometimes referred to as Domain Shift (Storkey, 2009). In our setting,

common reasons for this phenomenon are that farm management practices can vary as one moves

to new regions and that growing seasons can shift, shrink, or extend by a few days as one moves

to new climatic regimes. In addition, the peak value in the time series of a spectral band (or





vegetation index) can increase or decrease due to differing climatic or management (e.g.

fertilization rates) conditions between regions.

We develop an approach that corrects for a shift in the mean of the features between the training

set and test set. Shifts in the composition of crop types between regions can cause apparent

feature shifts, with the overall feature mean vector moving towards those of crops with increased

proportion in the new region and away from those of crops with decreasing proportion. Our

approach therefore decomposes shifts into those that arise from a change in distribution of the

test labels (i.e. apparent feature shifts) and those that arise from true feature shifts. In particular,

suppose we have $R$ regions and $K$ crop types, and for $r = 1, \dots, R$ and $k = 1, \dots, K$ let $\vec{\mu}_{r,k}$ denote

the mean feature vector in region $r$ for crops of type $k$. To adjust for translations in the mean of

the features between regions, we consider the following additive model:

$$\vec{\mu}_{r,k} = \vec{a}_r + \vec{b}_k \quad \text{for } r = 1, \dots, R \text{ and } k = 1, \dots, K$$

where $\vec{a}_r$ and $\vec{b}_k$ are constant vectors for region $r$ and crop type $k$. This additive model says that

the mean feature vector $\vec{\mu}_{r,k}$ of each region and crop type stratum can be written as the sum of a

regional component and a crop type component; the additive model assumes there are no

interactions between the crop type and region variables.

Suppose we have labeled data only for region $r_{train}$, but we know the distribution of the labels

for all regions. That is, if $p_{r,k}$ equals the proportion of labels in region $r$ that are of type $k$, we

know $p_{r,k}$ for $r = 1, \dots, R$ and $k = 1, \dots, K$. Then we can estimate the regional contributions $\vec{a}_r$

for $r \neq r_{train}$ with the following two-step procedure:





1) Let $\vec{a}_{r_{train}} = \vec{0}$ without loss of generality, and estimate each $\vec{b}_k$ with $\hat{\vec{b}}_k$ where $\hat{\vec{b}}_k$ is the sample mean vector of the features for crop type $k$ in the training region

2) Estimate each $\vec{a}_r$ with $\hat{\vec{a}}_r = \bar{X}_r - \sum_{k=1}^{K} p_{r,k} \hat{\vec{b}}_k$ where $\bar{X}_r$ is the sample mean vector of the features in region $r$. Repeat for each $r \neq r_{train}$.

Once completing the above steps, for each $r \neq r_{train}$, we subtract the estimated regional contributions $\hat{\vec{a}}_r$ from the feature vector of each test point in that region before applying a classifier. In particular, if you train a classifier $\hat{l}(\cdot)$ on region $r_{train}$ and wish to classify a point on region $r$ with feature vector $x$, our proposed Feature Shift Adjusted (FSA) classifier is given by $\hat{l}_{FSA}(x) = \hat{l}(x - \hat{\vec{a}}_r)$. In essence, this proposed classifier removes the estimated regional effects under the previously described additive model. Note that this FSA classifier can easily be combined with the PSA classifier $\hat{l}_{PSA}(\cdot)$ of Section 2.1 using $\hat{l}_{FPSA}(x) = \hat{l}_{PSA}(x - \hat{\vec{a}}_r)$ where $r$ is the test region. We will call this the combined Feature and Prior Shift Adjusted (FPSA) classifier, and it is recommended to combine the two approaches in settings where there is both substantial feature shift and prior shift.

While the additive model is unlikely to hold exactly in realistic settings, there are a few reasons that justify our proposed use of an additive model. First, if the interaction terms between the crop type and region that contribute to $\vec{\mu}_{r,k}$ are sufficiently small, the FSA classifier will remove most of the regional effects and will still confer significant benefits. Second, in settings where labeled data is only available in one region $r_{train}$ and the remaining regions have aggregate-level crop statistics but do not have available field-level labels, alternative models that include interaction





terms between the crop type and region, such as a multiplicative one with main effects for crop type and region, cannot be fit; only models without interactions can be fit. Ultimately, it is better to correct for main effects with a method such as FSA or FPSA when possible than to completely ignore the main effects, even when there are large interaction terms. Third, the use of models with assumptions that do not necessarily hold are routinely justified empirically by checking whether they lead to more accurate predictions. To this end, in Section 4 we investigate the accuracy of FPSA when compared to alternative transfer learning methods.

## 3) Data and methods

### 3.1) Field-level polygons and labels

### 3.1.1) Field-level dataset from France

The French Land Parcel Identification System is a very large and accurate dataset of parcels in France with crop type labels exclusively obtained by on the ground field surveys. From this dataset, we randomly sampled 100,000 labeled parcel centroids that were recorded in 2017 and were within the rectangular region between 42.3º N and 45.1º N and between 0.4º W and 4.9º E (which completely encompasses the Occitanie region, depicted in the top panel of Figure 1). Of these labeled parcels, 74,169 fell within Occitanie and were included in our analyses.





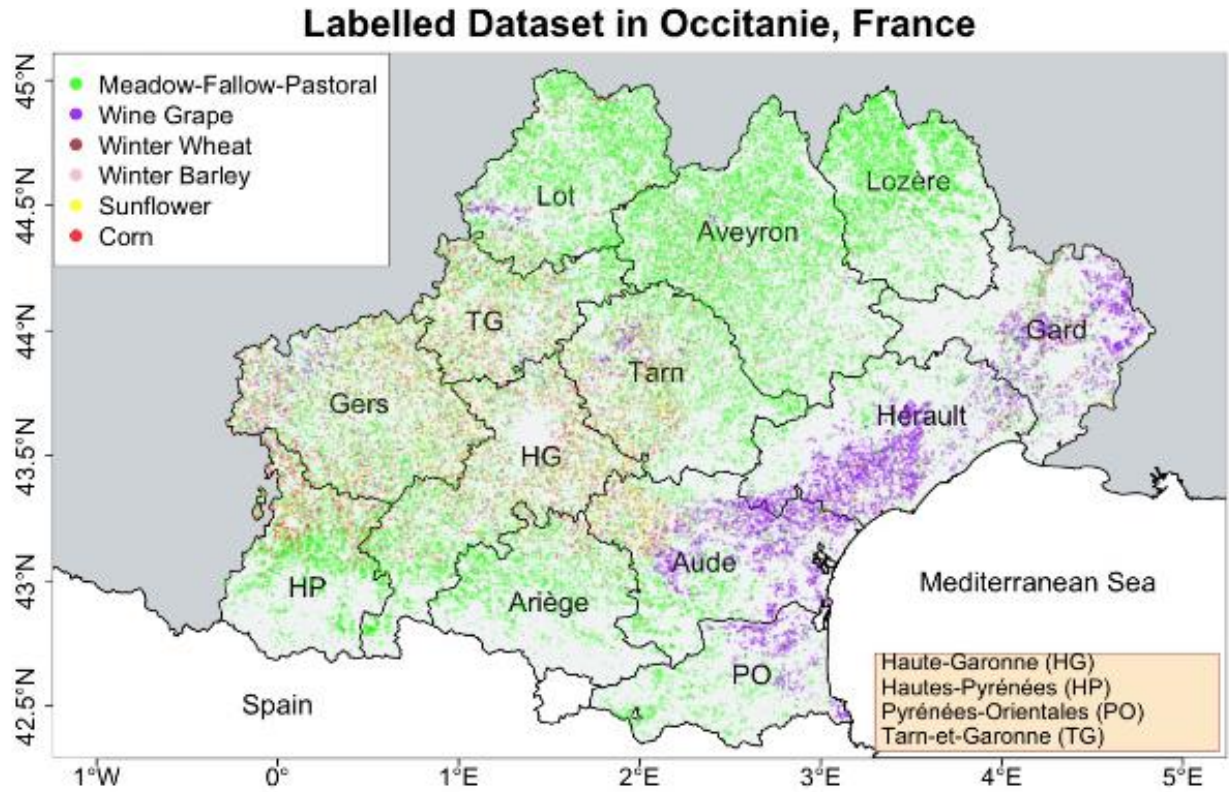

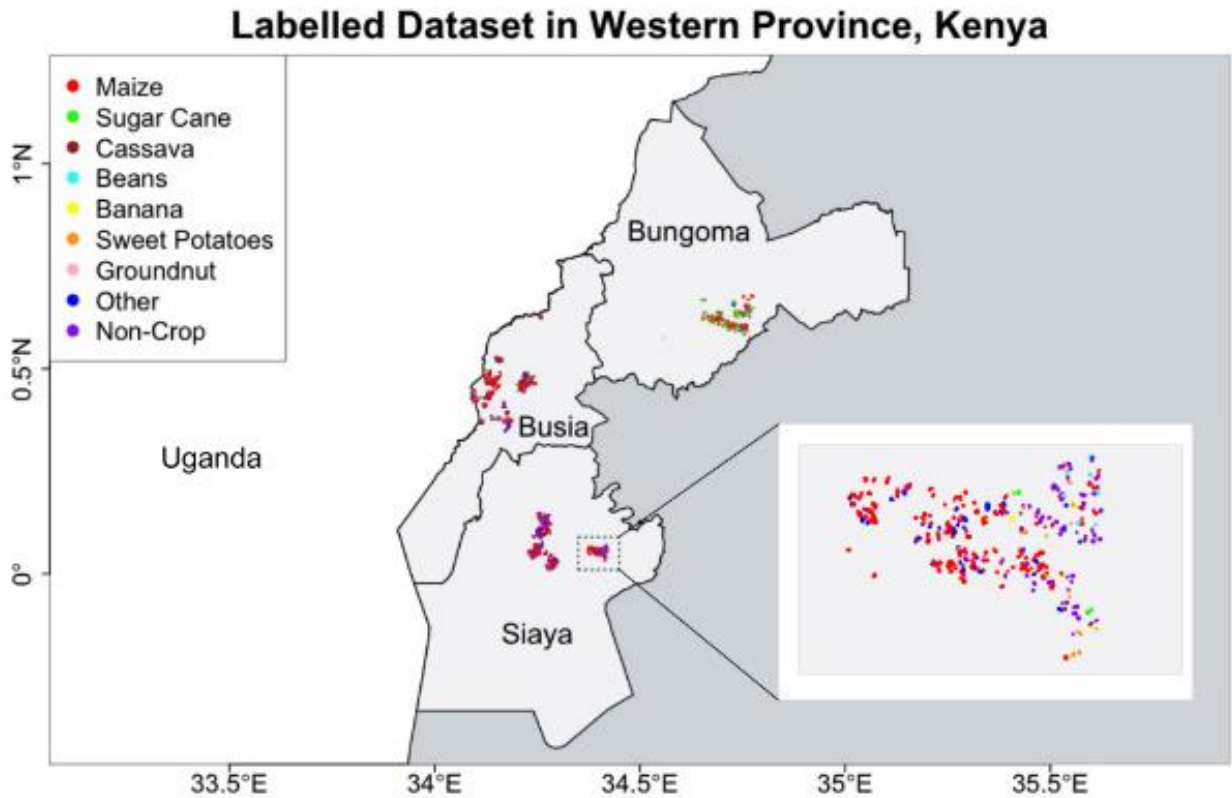





Figure 1: Top: Map of the departments in Occitanie, France. Our dataset includes labeled data from all 13 departments in Occitanie. Parts of France that lie outside the Occitanie border are colored in dark grey. For departments with abbreviated names, the orange box includes the full name of the department along with the abbreviation in parenthesis. The colored points on the plot depict the spatial distributions of the parcels in our simplified six-class, classification dataset in Occitanie, France, colored by crop type. Bottom: Map of 3 regions with Western Province, Kenya. Our dataset includes labeled data from the regions Bungoma, Busia, and Siaya (colored in light grey). Parts of Kenya that lie outside of these three regions are colored in dark grey. The colored points on the plot depict spatial distributions of the labeled pixels in our nine-class, classification dataset in Western Province, Kenya, colored by crop type.

To simplify our classification task, we identified the 17 most common field labels among the 74,169 sampled parcels and removed 12,892 parcels that had labels which were not among these 17. We then removed 8,613 parcels that had labels of either "field border", "buffer strip" or "agricultural area not exploited." This led to a final dataset of 52,664 fields with 14 different classes. We further consolidated the 14 remaining classes into 6 classes by changing the label of both "soft winter wheat" and "hard winter wheat" to "winter wheat", and by assigning many different types of meadow, fallow and pastoral lands the label meadow-fallow-pastoral (MFP). See Appendix B for a table of the field labels and counts in the original subsample of 74,169 parcels from Occitanie as well as a color-coded depiction of how we simplified our classification task. In subsequent data processing steps, an additional 23 parcels were removed due to failure to fit a harmonic regression to each spectral band caused by too many days with cloud cover detected (see Section 3.4.2 and Section 3.4.3). In summary, because we were focused on discriminating between common crop types as well as MFP lands, we simplified our classification task by using the labels from the French Land Parcel Identification System to remove both uncommon classes and non-cropped and non-MFP areas. As a result, our training and validation datasets included six different classes, and did not include an "other crops" class.





See Figure 2 for a histogram of the labels in each department from our post-processed and simplified dataset of 52,641 parcels. Also see the top panel of Figure 1 for the spatial distribution of the labels in our post-processed and simplified dataset.

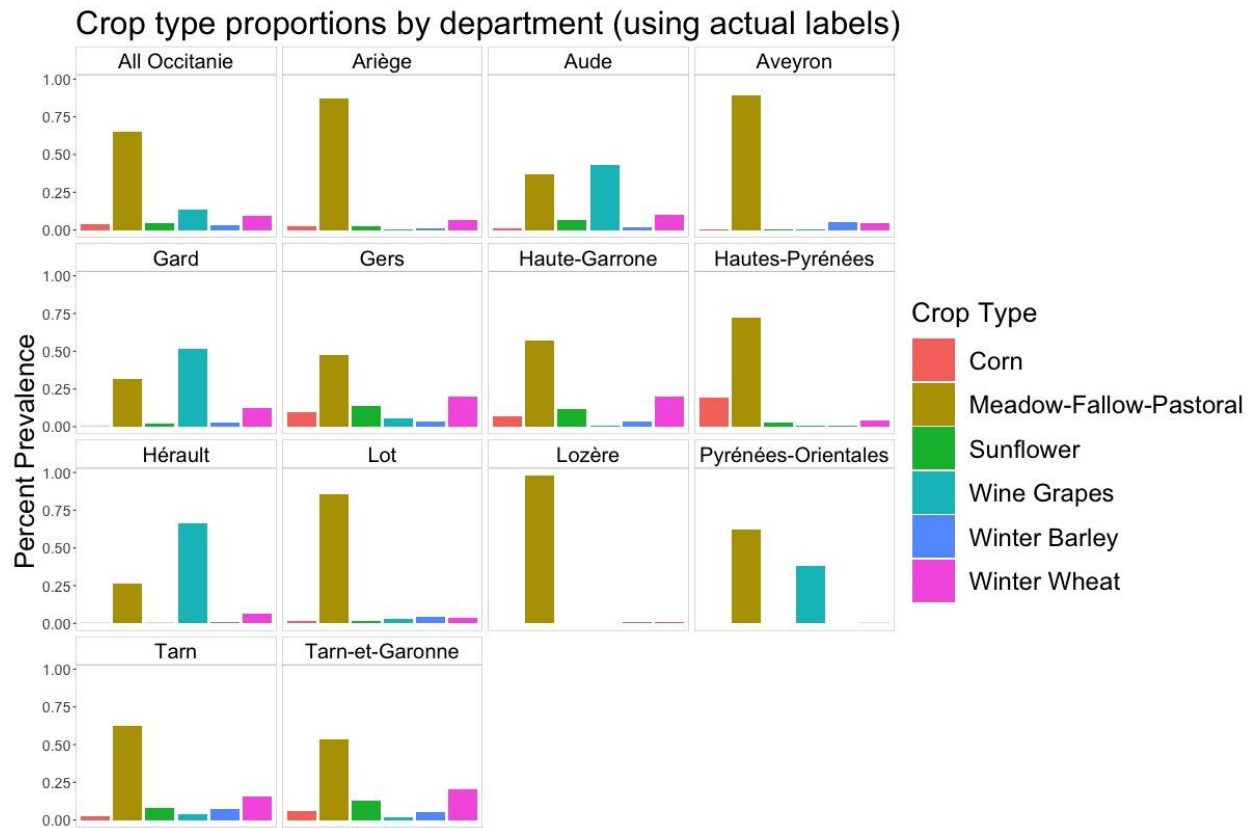

Figure 2: Proportion of crop types with each label in our processed dataset of 52,641 field centroids from Occitanie, France. The top left barplot displays the estimated percent prevalence of each crop type across our entire dataset while the other 13 barplots give department specific crop label distributions.

### 3.1.2 Field-level dataset from Kenya

In Western Province, Kenya, we utilize a dataset on field boundaries for 16,964 fields during the long-rains season (Mar-Sep) in 2017 from Jin *et al*., 2019. These data span three distinct districts in the province (see bottom panel of Figure 1), which differ both in crop mix (Figure 3) and





climatic and soil conditions. We removed all fields that did not have a crop type label and

further removed all 92 fields belonging to the small pilot dataset that was gathered to check

measurement instrumentation, leaving 8,041 fields. Because the Kenya dataset had much fewer

labeled fields than the France dataset, to increase our sample size in Kenya, instead of

constraining our attention exclusively to the field centroids, we used Google Earth Engine to

identify all pixels within each labeled field that were buffered by at least 5 meters away from the

field boundary. Each buffered pixel was considered a separate training or test point in our

classification task.

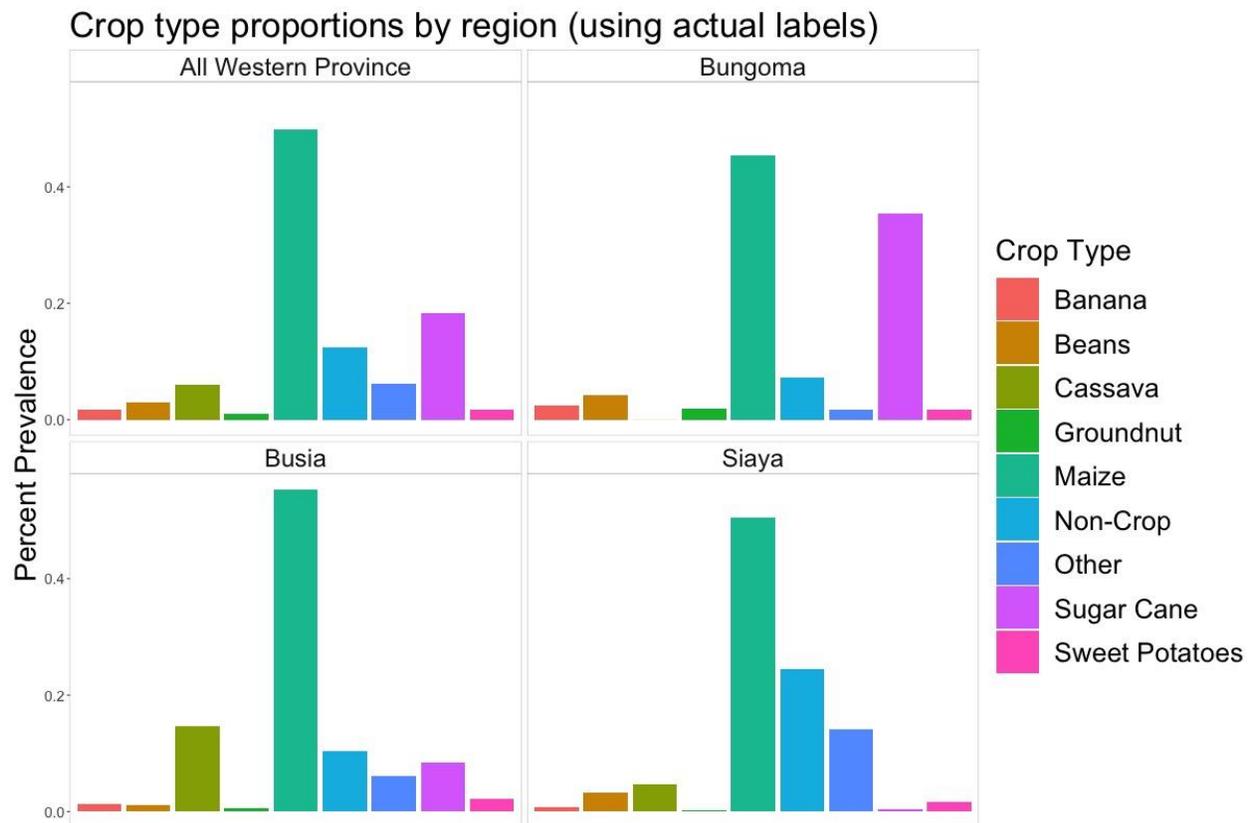

Figure 3: Proportion of crop types with each label in our processed dataset of 39,881 buffered pixels from Western Province, Kenya. The top left barplot displays the estimated percent prevalence of each crop type across our entire dataset while the other 3 barplots give region specific crop label distributions.





### 3.2) Aggregate-level data

### 3.2.1) Aggregate-level data in France

To estimate the proportions of each label within each department of Occitanie, we used official department-level crop statistics (French Ministry of Agriculture and Food, Agreste, n.d.). In particular, we extracted information from a Land Use dataset, Cultivated Crops dataset, Grape Production dataset, and a Forage and Meadow dataset. For each department in Occitanie for the year 2017, we extracted the estimated developed area for each of our six classes using the dataset entries described in Table A2 in Appendix C.

Because the data from France that was used for testing our methods included just one centroid pixel per parcel rather than all the pixels in each parcel, in order for PSA or FPSA to be suitable for application to our testing sets, we needed an estimate of the proportion of parcels of each crop type within each department of Occitanie. Therefore, we had to transform the information about the total area of each of our six crop type classes taken from official department-level crop statistics into department-level estimates of the proportions of parcels of each crop type. To do this, we divided the aggregate area of a given crop in a given department by the average area of a field for that crop type and that department. While these department-level field area averages were calculated using field-level data (from the French Land Parcel Indentification System in this case), which would not be available for the proposed use cases of our methods, such information would not be needed if we were applying a transferred classifier to all of the agricultural pixels in a new region. This is because if we were to apply PSA or FPSA to all of the agricultural pixels in a new region, prior probabilities that a given pixel is of each class will be proportional to that class's total crop area in the new region, which can be estimated from government statistics





without access to field-level data. Our need to transform total crop area estimates to parcel crop type proportions is merely an artifact of the fact that we sought to test our classifiers on parcel centroids rather than on all agricultural pixels in a new region and, therefore, does not raise concerns about the practical applicability of our proposed methods.

By comparing Figures A1 and A2 in Appendix C, we see that estimates of crop composition using aggregate-level data are quite accurate. The aggregate-based estimate was especially good for estimating the relative abundance of the non-MFP classes (see Appendix C for a discussion of why the aggregate-based estimates for the proportion of points with the MFP label were slightly less accurate than those for the relative abundance of the non-MFP classes).

### 3.2.2) Aggregate-level data in Kenya

At the time of this writing, aggregate-level regional data from Kenya for 2017 for the suite of crop types we considered were unavailable. Instead, as a stand-in, we calculated the aggregate distribution of the crop type labels in each region from the actual field-level labels.

### 3.3) Sentinel-2 multi-temporal satellite imagery

We used time series of optical imagery collected by the Sentinel-2 constellation to classify crop types, as the imagery is both publicly available and at high enough resolution (10-60 m) to map smallholder fields in Western Province, Kenya. In particular, we extracted the multispectral time series measurements from 2017 taken by Sentinel-2 of the parcel centroids in our labeled dataset from France and of the buffered pixels in our labeled dataset from Kenya using Google Earth





Engine. The Sentinel-2 satellites have a 5-day revisit period and capture thirteen spectral bands ranging from blue to short-wave infrared. While the spectral band resolutions range from 10-60 meters, all bands were resampled to 10 meter resolution using the Google Earth Engine default of nearest neighbor interpolation. Temporally, all satellite observations extracted from Western Province, Kenya occurred on the same 50 days throughout 2017, with the time between observations alternating aperiodically between 5 and 10 days. Meanwhile, the days in the time series were not consistent throughout Occitanie in 2017, due to the region spanning multiple Sentinel-2 tracks; however, on average the time between two sequential observations of the same parcel was 4.6 days. Cloudy observations were still extracted along with outputs of cloud detection classifiers, so that cloudy observations could be removed in a later step as described in Section 3.4.2.

The European Space Agency (ESA) distributes a top-of-atmosphere reflectance product (Level-1C) as well as a surface reflectance product (Level-2A) that can be derived using a toolbox provided by ESA. At the time of writing, ready-to-use Level-2A imagery was not available over Kenya in the year 2017. As a result, we used the Level-1C product in all following analyses. There is evidence that land cover classification using top-of-atmosphere products and surface reflectance products are comparable, since classification is concerned with relative spectral differences (Rumora et al., 2019; Rumora et al., 2020); however, since our study area spans multiple Sentinel-2 tracks and different atmospheric corrections, the relative spectral differences could still vary across the study region. This suggests that using Level-2A imagery would have led to higher classification accuracies than the ones observed in our results.





### 3.4) Preprocessing steps

We implemented three pre-processing steps to turn each year-long multispectral time series into a 70-dimensional feature vector, as described below.

### 3.4.1) Green Chlorophyll Vegetation Index

We computed the Green Chlorophyll Vegetation Index (GCVI) using the formula GCVI=NIR/GREEN-1. The index was designed to capture chlorophyll content in crops (Gitelson *et al.*, 2005) and has been shown to be an important feature in crop classification settings (Wang *et al.*, 2019).

### 3.4.2) Removing points with cloud cover and measurement issues

While extracting the Sentinel-2 data from Google Earth Engine, we also extracted the QA60 value and the Hollstein Quality Assessment measure for each parcel centroid in Occitanie and for each buffered pixel in Western Province, Kenya to help us identify measurements that were disrupted by cloud cover or other issues. The QA60 measurement labels each satellite pass for a given pixel as either having no clouds, cirrus clouds or dense clouds, by using a decision tree with measurements from the blue, cirrus, SWIR1 and SWIR2 bands as inputs followed by morphology-based operations to avoid isolated pixels being labeled as no clouds, cirrus, or dense (European Space Agency (ESA) Technical Guides, n.d.). The Hollstein Quality Assessment (Hollstein *et al.*, 2016) uses a more complex decision tree to determine whether a given pixel at a certain time is "clear" or if it has cloud, snow, water, or shadow cover. For each centroid in France and pixel in Kenya, we only retained points in the time series that were considered clear and cloud-free by both the QA60 value and the Hollstein Quality Assessment value. All time





series points that did not meet these criteria were removed. We used both quality assessments in order to be conservative, as removing a few extra points in each time series would not skew our fitted harmonic regression coefficients whereas retaining points with corrupted measurements can lead to inaccurate fitted harmonic regression coefficients.

### 3.4.3) Harmonic regression

After removing all time series points that had some indication of cloud cover or other issues, we transformed our multispectral time series into harmonic features that capture crop phenology using harmonic regression. This approach has been shown to be useful in characterizing and classifying crop types (Jakubauskas *et al.*, 2002; Wang *et al.*, 2019) and has the advantage of constructing consistent features despite inconsistent cloud cover. Our harmonic regressions reduced the time series of each spectral band (or vegetation index) into 5 harmonic features because we set n = 2, where n the number of different frequencies used to construct the Fourier basis in the regression. This choice was determined to be optimal in the crop classification setting of Wang *et al.*, 2019. To implement harmonic regression, we solved the following regression problem for each filtered time series.

$$Y_{i,j,t} = c_{i,j} + \sum_{k=1}^{2} [a_{i,j,k} \cos (2\pi kt) + b_{i,j,k} \sin (2\pi kt)] + \varepsilon_{i,j,t}$$

Where $Y_{i,j,t}$ is the measurement at time $t$ for band $j$ in pixel $i$, where $c_{i,j}$, $a_{i,j,1}$, $b_{i,j,1}$, $a_{i,j,2}$ and $b_{i,j,2}$ are the 5 harmonic coefficients for pixel $i$ and band $j$, and where $\varepsilon_{i,j,t}$ are error terms with mean zero that are assumed to be independent and identically distributed with mean 0, and $t$ is the time variable in years which is set to be zero on April 1st. Note that while $c_{i,j}$, $a_{i,j,1}$, $b_{i,j,1}$, $a_{i,j,2}$ and $b_{i,j,2}$ cannot be known exactly with only 50-100 measurements per pixel per band per





year, it can be estimated using linear regression onto the harmonic basis given in the above regression problem.

After fitting linear regression, our estimates for the 5 coefficients for band $j$ in pixel $i$, $\hat{c}_{i,j}, \hat{a}_{i,j,1}, \hat{b}_{i,j,1}, \hat{a}_{i,j,2}$ and $\hat{b}_{i,j,2}$ are used as features for our classification task. Note that for each pixel $i$, we implement this harmonic regression for 13 different spectral band time series, as well as for a 14th GCVI time series. Since each such regression yields five features, we have $14 \times 5 = 70$ harmonic features for each pixel. As such, we reduce each pixel's features from a multispectral time series to 70 harmonic features.

It should be noted that when a time series for a given pixel has fewer than 5 distinct time values, the above regression cannot be implemented. This can occasionally occur if a majority of the days for a pixel are deemed to not have clear measurements according to the QA60 mask or the Hollstein Quality Assessment. This occurred in France for 23 out of our 52,664 parcel centroids and in Kenya for 0 out of our 39,881 buffered pixels.

### 3.5) Base classifiers used

While the methodology in Section 2 can be applied to a broad set of classifiers, we focused on applying it to the Linear Discriminant Analysis (LDA) classifier and the Random Forest classifier. LDA is a classical classification method that choses "optimal" boundaries under a Gaussian mixture model assumption and predicts the labels of newly sampled points based on which side of the boundary the points lie. The LDA classifier first fits a multivariate Gaussian mixture model with common covariance to the labeled points in the training data and then





computes the Bayes optimal classification rule for unlabeled data drawn from this fitted mixture

model. Furthermore, the LDA classifier will always have linear decision boundaries in feature

space (Hastie *et al.*, 2009). Notable to our discussion of prior shift and feature shift adjustment is

that the LDA classifier first calculates the proportion of labels of each class to estimate the

Gaussian mixture weights as well as the sample mean vector of the features for each class to

estimate the Gaussian mixture mean vectors. The PSA method (see Section 2.1) will adjust the

trained LDA classifier so that it uses the correct mixture weights on the test set, while the FSA

method (see Section 2.2) will adjust the classifier so that it uses the appropriate feature mean

vectors for each class on the test set as parameter vectors for the assumed Gaussian mixture

model. Therefore, our FPSA method is well suited for correcting an LDA classifier.

Random Forest (Breiman, 2001) is a widely-implemented, state-of-the-art classifier that is

popular in crop classification settings (Waldner *et al.*, 2017; Wang *et al.*, 2019). The Random

Forest classifier works by aggregating the votes of trees trained on different bootstrapped

samples of the data. To add additional randomness to the training process, when each decision

tree split is made, only a random subset of the features is considered (Hastie *et al.*, 2009). The

posterior probabilities returned by a Random Forest classifier are simply the proportion of trees

that vote for each class; unlike those of the LDA classifier, these cannot be interpreted as

probabilities under a certain set of assumptions. The PSA method (see Section 2.1) will reweight

the pseudo-probabilities (i.e. the proportion of trees voting for each class) on the test set by the

ratio of that class's prevalence on the test set over its prevalence on the training set and will pick

the class with the highest reweighted pseudo-probability as the fitted label. Meanwhile the FSA

method (see Section 2.2) will translate the features on the testing set in order to maintain

consistent class-wise feature mean vectors on the training and test set before the fitted Random





Forest is applied to test set points. Note that unlike for LDA, decision boundaries for a Random Forest do not necessarily correspond to distances in feature space from each of the class-wise feature means. Therefore, our FPSA method is not perfectly suited for correcting a Random Forest but can be useful as long as the pseudo-probabilities are not egregiously inaccurate and as long as the Random Forest decision boundaries are not irregular enough for mean feature shift adjustment to cause harm.

### 3.6) The validation procedure: Train on one, test on rest

For the transfer approaches that we consider (with the exception of the Oracle described in Section 3.9), we conduct the following experiment to validate its accuracy in Occitanie, France. We split our dataset with labeled data from Occitanie into 13 labeled datasets (one for each department). We then designate one of those 13 single-department datasets as the training set and the remaining 12 single-department datasets are considered separate test datasets. Then, for each of the 12 departments outside of the training department, we train a classifier on the designated training set that is geared towards transferring to that target department (leveraging the mean feature vector on the target department or the prior information about the crop type distribution on the target department or both), apply it to the testing dataset from that target department with held out labels, and store the estimated labels. After repeating this procedure for all 12 departments outside of the training department, we compare estimated labels with the held out labels from the 12 testing datasets and store the results aggregated across the 12 testing departments in a confusion matrix. The overall accuracy as well as the class-wise producer's accuracies, user's accuracies, and $F_1$-scores are then calculated using this confusion matrix. This experiment and these accuracy metrics give insight into how well a particular transfer approach





works when there is labeled data in one particular department of Occitanie and we want to classify crops in the remaining departments of Occitanie. To give a broader picture of the effectiveness of a particular transfer approach, we repeat the experiment for 11 different possible choices of training departments. We do not conduct the experiment with Lozère or Pyrénées Orientales as the designated training department because these departments each had at least one of the six crop types absent from the labeled dataset, making it impossible to train a classifier on these regions that can distinguish between all six classes; however, we still consider Lozère and Pyrénées Orientales when testing our classifiers.

In Kenya, we conduct similar experiments to validate the accuracy of the different transfer approaches that we consider. For each experiment, instead of splitting the labeled dataset into one single-department training dataset and 12 single-department testing dataset, we split the labeled dataset into one single-region training dataset and two single-region testing datasets. For example, when Busia is the designated training region, we train a classifier in Busia geared for being transferred to Bungoma, compute the fitted labels in Bungoma, and then do the same for a classifier trained in Busia geared towards being transferred to Siaya. Using the held out true labels in Bungoma and Siaya and the fitted labels we make a confusion matrix aggregating the results in these two testing regions and compute various accuracy metrics for the classifier. We conducted these experiment in France and Kenya for multiple transfer learning approaches using both LDA and Random Forest as a base classifier to compare the effectiveness of different transfer learning approaches. The multiple transfer learning approaches that we consider are outlined in Table 1 and described in Sections 3.7-3.10 in more detail.





### 3.7) The main three transfer approaches

Our first main approach, which we call Unadjusted Transfer (UAT entry in Table 1), was the naïve approach where we simply trained the classifier in one region and applied it to the target regions without using any prior shift or feature shift adjustments.

Our second approach (PSA entry in Table 1) was to split up the test set by the 12 test departments (in France) or 2 test regions (in Kenya), and for each test department or region, we applied the PSA classifier described in Section 2.1. After determining the fitted labels in each of the testing regions using the PSA classifier, we computed and stored the confusion matrix aggregated all the test departments or regions.

Our third approach (FPSA entry in Table 1) was to again split up the test set by departments or regions and apply the FPSA classifier described in Section 2.2. After determining the fitted labels in each of the testing regions using this classifier, we computed and stored the confusion matrix aggregated all the test departments or regions.

**Table 1:** A summary of the classifier adjustment methods we considered. The first two methods listed are for baseline comparisons and do not address either prior shift or feature transformations between regions. The last method listed is unusable in practice but gives us an estimated upper bound on the accuracy of transfer learning methods when the given base classifier is used.

| Acronym | Depiction in Section 4 barcharts | Description | No adjustment (Neither), prior shift adjustment only, or both prior and feature shift adjustments (Both) | Validated using train on one, test on rest procedure described in Section 3.6? | Described in Section |
|---|---|---|---|---|---|
| GMC | Dark Grey Bar | Only guesses the major class. In Occitanie, only meadow-fallow-pastoral is guessed while in Western Province, Kenya only maize is guessed. | Neither | NA (this approach does not require training a classifier) | NA |
| UAT | Light Grey Bar | Transfers classifier without any adjustments | Neither | Yes | 3.7 (briefly) |
| PSA | Yellow Bar | Adjusts for prior shift by reweighting the classifier's posterior probabilities | Prior shift adjustment only | Yes | 2.1, 3.7 |





| FPSA | Orange Bar | Adjusts for feature translation by correcting for mean feature shifts between regions (and, in addition, uses prior shift adjustment) | Both | Yes | 2.2, 3.7 |
|---|---|---|---|---|---|
| SMOTE-PSA | Solid Black line (Column 3) | Artificially rebalances the training set, to have the same crop type composition as the test set using SMOTE | Prior shift adjustment only | Yes | 3.8.1 |
| zT-FPSA | Dashed Black line (Column 4) | z-transforms the features in each region then applies a prior shift adjustment | Both | Yes | 3.8.2 |
| zT-SMOTE-FPSA | Solid Black line (Column 4) | Artificially rebalances the training to have the same crop type composition as the test set using SMOTE, then applies z-transformation | Both | Yes | 3.8.2 |
| Oracle | Light Blue Bar | Uses Data from each target region to train a classifier for that target region. Accuracy is computed with 10-fold cross-validation | Neither | No | 3.9 |

## 3.8) Alternative shift adjustment approaches

We also compared our methods to simple approaches that may be initially considered for adjusting for prior shift and feature shift. First, we describe an alternative prior shift adjustment approach that uses data augmentation to balance the training data such that the augmented training data has the same crop type composition as the target region. Then, we describe two alternative feature shift adjustment approaches that involve centering and standardizing the features in each region, and one of these approaches also relies on dataset balancing. A comparison of our method to the alternative approaches described in this section can be found in Section 4 and can also be seen by looking at the solid and dashed lines in Figures 6, 9, A3 and A4.





### 3.8.1) Alternative prior shift adjustment approaches

A natural prior shift adjustment approach to consider is to simply sample points on the training region so that the sample has the same crop type distribution as the target region. The goal though is ultimately to train a classifier that will transfer well to multiple nearby regions, each with different crop type compositions, so it would be impractical to calibrate the crop type distribution of the collected labels from field-surveys to just one of the nearby regions. That being said, once the labeled data for the training set is collected, the training set data could be synthetically rebalanced as needed to have any desired distribution of the labels. Therefore, a common approach for prior shift adjustment is to first artificially rebalance the training set so that it has the same label distribution as the test set and to then fit a classifier to the artificially rebalanced dataset. There are a number of ways to artificially rebalance the training data, some of which have been considered for remote sensing applications (Waldner *et al.*, 2019). Examples include undersampling points or oversampling points of each class in the training set until an artificially rebalanced training set with the same label distribution as the test set is achieved. For oversampling, the simplest approach is to randomly choose points to replicate. An alternative and popular oversampling method is the Synthetic Minority Oversampling Technique (SMOTE) which generates new points by randomly picking a point, then picking one of the k nearest neighbors of that point (we used k=5 as it was the default value of the `SmoteClassif` function in the `UBL` R package (Branco *et al.*, 2016), which we used for implementation), and finally generates a synthetic instance along the line segment connecting the two randomly selected points. In (Waldner *et al.*, 2017) SMOTE was used in order to correct for prior shift in a binary crop classification setting and was shown to lead to significant increases in classification accuracy.





Therefore, we compared the PSA method presented in Section 2.1 to SMOTE-based prior shift adjustment (which we denote with the acronym SMOTE-PSA). As described in Section 3.6, after choosing a designated training region, we recalibrated and reapplied SMOTE-PSA to each of the different target testing regions separately, and then aggregated the confusion matrix and accuracy across the different testing regions.

### 3.8.2) Alternative feature shift adjustment approaches

Another approach for feature shift adjustment is applying a z-transformation to both the testing and training set. To implement a z-transformation one subtracts the mean and divides by the standard deviation of each feature among the points in that set. By way of comparison to our feature shift adjustment methodology, in each department in Occitanie and in each region in Western Province, Kenya we applied a z-transformation to our features. After implementing the z-transformations, we repeated our PSA approach described in Section 3.7, which is based on the prior shift adjustment method presented in Section 2.1. This approach is included in Table 1 with the acronym zT-FPSA.

We did not expect the standard z-transformation to lead to good performance in settings where the test set label distribution is very different than the training set label distribution. This is because when a feature has a different mean or variance between two different regions, the





differing mean and variance can be largely driven by having different class label proportions

rather than a systemic transformation of the features between regions. Ideally, we want to remove

the effects of systemic feature transformations between regions but without removing variations

in the features that are due to differing class labels (which are essential to high-performance

classification). Therefore, we also considered using a z-transformation but only after artificially

rebalancing the training dataset. In particular, first we used SMOTE to artificially rebalance the

training set label distribution to match that of the testing set, and then we applied a z-

transformation to both the artificially rebalanced training set and to the test set. Because the

training set was rebalanced to match the label distribution on the test set, we did not use the

method presented in Section 2.1 to adjust for label distribution shifts. As described in Section

3.6, after choosing a designated training region, we reapplied the z-transformation and SMOTE-

based feature shift adjustment method for each of the different target testing regions separately,

and then aggregated confusion matrix and accuracy across the different testing regions. This

approach is included in Table 1 with the acronym zT-SMOTE-FPSA.

### 3.9) The Oracle Classifier

In order to assess the gap between the accuracy of our transfer learning approaches and the best

accuracy they can achieve with the same features and base classifier model, we also considered

an Oracle classifier that is trained using the test set labels. To compute the accuracy of an Oracle

Classifier for either LDA or Random Forest, for each region, we computed the 10-fold cross-

validation accuracy when the training and test sets were random subsets of that region (in Kenya,

when performing 10-fold cross-validation, the random subset assignment ensured that no plot of





land would have pixels in both the training and testing set, but this precaution was not necessary in France where each plot of land only had one pixel in our post-processed dataset. This precaution was also not necessary in Kenya when validating the transfer approaches described in Sections 3.7 and 3.8 because the training and testing sets had pixels from different polygons by default, as the training and testing pixels were from entirely different regions). The Oracle classifier's accuracy for a particular "training" region was simply the weighted average of the cross-validated accuracies of the other regions in the dataset. These weighted averages give a good estimate for an upper bound on the accuracy that can be achieved by transfer learning adjustment approaches when training on a particular region and using a particular base classifier. It is important to note that such a classifier cannot be used in practice when the labels on the target regions are unavailable, and because it would not be used in practice, it is the only classifier whose accuracy we did not assess using the validation procedure presented in Section 3.6.

## 4) Results

In Sections 4.1-4.4 we describe in detail our data visualizations and results from Occitanie, France in order to familiarize the reader with the analyses and the presentation of their results. In Section 4.5 we show and briefly discuss our results from Western Province, Kenya.

### 4.1) Data visualizations

We first visualize an example of feature shift between regions to demonstrate that the same crop type can have different average profiles in different regions. Figure 4 compares the time series of





the GCVI values averaged over all corn, sunflower, winter barley and winter wheat data points in

Aude and in Tarn. To smooth the time series, GCVI averages were taken over 15-day bins after

excluding observations with cloud cover or other quality issues (see Section 3.4.2). We can see

from Figure 4 that crops tend to have a larger peak GCVI value in Tarn than they do in Aude, as

well as larger GCVI values outside of the growing season. The particular variation in mean

GCVI time series observed between Aude and Tarn is consistent with an additive signal

transformation, which would correspond to a translation of our harmonic features. Appendix D

defines the Fixed Signal Addition transformation and describes how this transformation affects

the harmonic features.

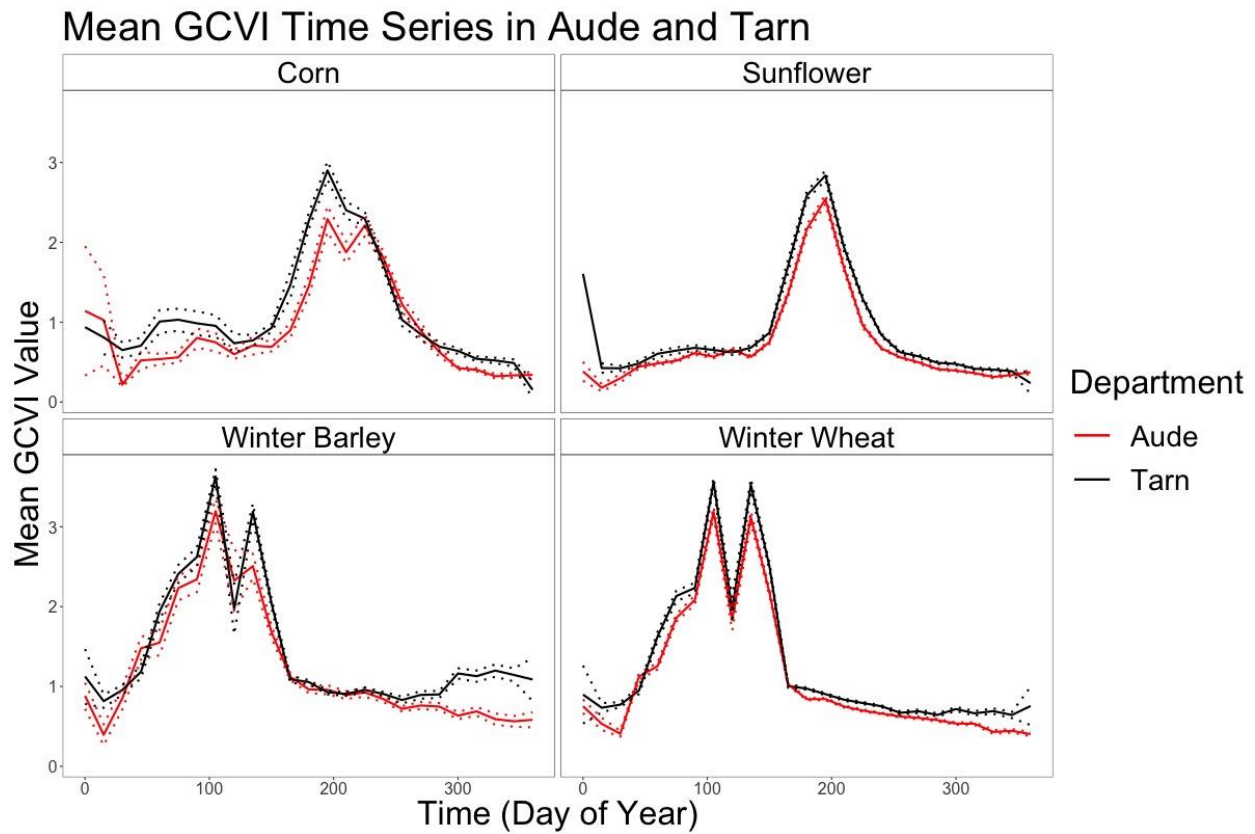

Figure 4: Plots of the GCVI time series averaged over all centroid pixels in Aude versus in Tarn for corn, sunflower, winter barley, and winter wheat. Sentinel-2 acquisitions that were not deemed clear were excluded. We binned time





into 15-day chunks before plotting averages over all clear observations within each bin. There are no gaps in any of the plotted time series, indicating that in both Aude and Tarn, for each plotted crop type, there is at least one parcel centroid with a clear observation in each 15-day bin. The dotted lines give 95% confidence intervals for the mean GCVI value, although temporal correlations are ignored when defining these confidence intervals (so we should not expect 95% coverage across the time series). It is clear from these plots that for the same crop type, GCVI values in Tarn tend to be higher than those in Aude.

To examine the shifts in the entire 70-dimensional feature space, we use t-stochastic neighbor embedding (van der Maaten and Hinton, 2008) to visualize the points in a two-dimensional representation (Figure 5). T-stochastic neighbor embedding (t-SNE) preserves neighborhoods, so points that are nearby in the 70-dimensional harmonic feature space appear nearby in the two-dimensional t-SNE representation (although the converse is not necessarily true). The t-SNE plots for the Occitanie dataset suggest that six classes could be reasonably well separated, making classifiers with high accuracies possible. One can also see that the distributions of t-SNE coordinates for a given crop type differ slightly between regions. This is clearest for the MFP class. Note that when running the t-SNE algorithm we used a scalar multiple of the first two principal component scores as our initial embedding rather than the default random initialization (see Kobak and Berens, 2019 for a description of this technique and a discussion of its advantages).





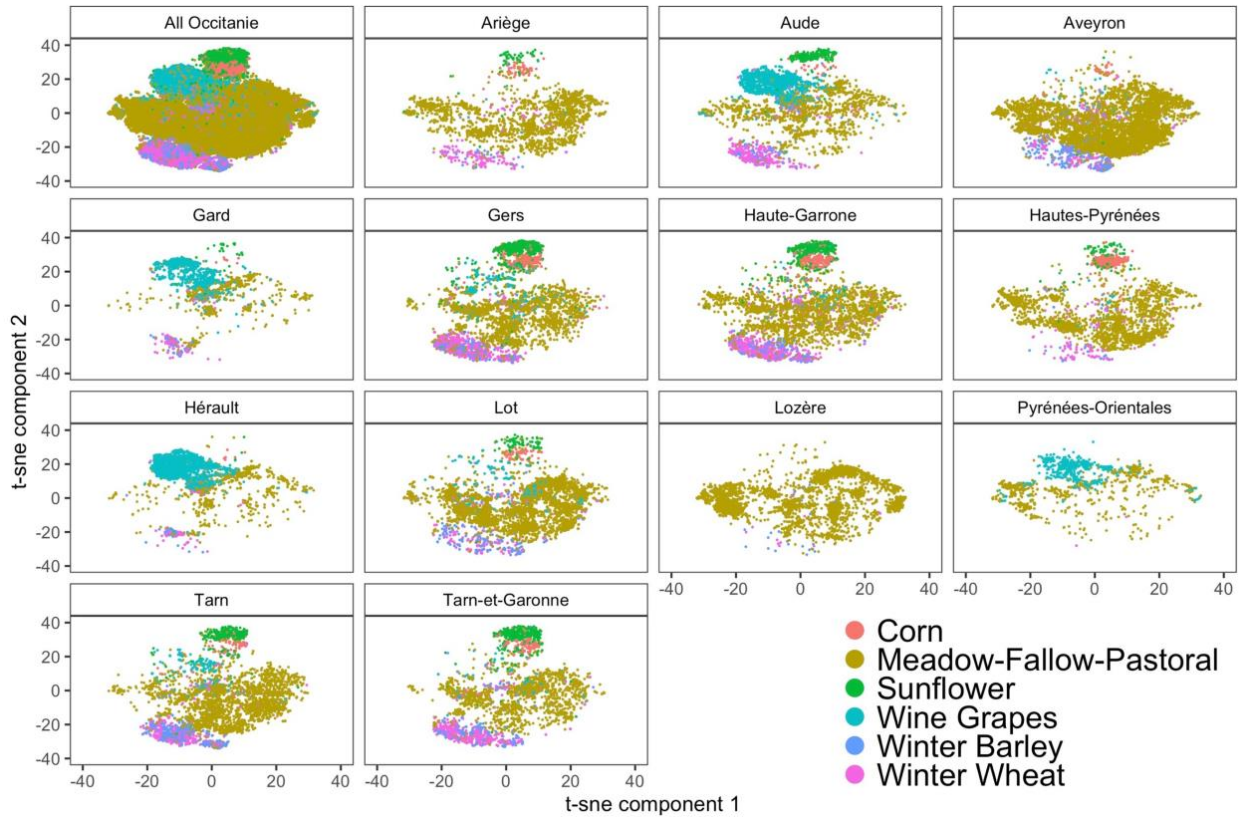

Figure 5: T-SNE plots of the harmonic features in Occitanie, France. The top left panel shows the t-SNE embedding of all 52,641 the points within Occitanie. The remaining panels show only the points within a particular department. Points that are near each other in the 70-dimensional harmonic feature space will appear close to each other in these t-SNE plots; however, points that are close to each other in these t-SNE plots are not guaranteed to be close to each other in the 70-dimensional feature space. This means that even though these plots may make it seem that distinguishing between winter barley and winter wheat would be difficult, it is possible that winter barley and winter wheat are actually well separated in our 70-dimensional harmonic feature space.

## 4.2) Classification results in Occitanie, France

### 4.2.1) With Linear Discriminant Analysis

The results for the 11 experiments in France, following the testing procedure described in Section 3.6, are plotted in Figure 6 (left panel). For reference we added the overall accuracy of the trivial classifier that just guesses the most common class (meadow-fallow-pastoral). In these





experiments, our base classifier was LDA. In each of the 11 experiments, we found that using the

FPSA approach led to an increase in overall accuracy ranging from 0.003 to 0.097 (mean=0.034)

when compared to the UAT approach. Further, in the 11 experiments, the FPSA approach led to

an increase in overall accuracy ranging from -0.002 to 0.083 (mean=0.020) when compared to

the PSA approach.





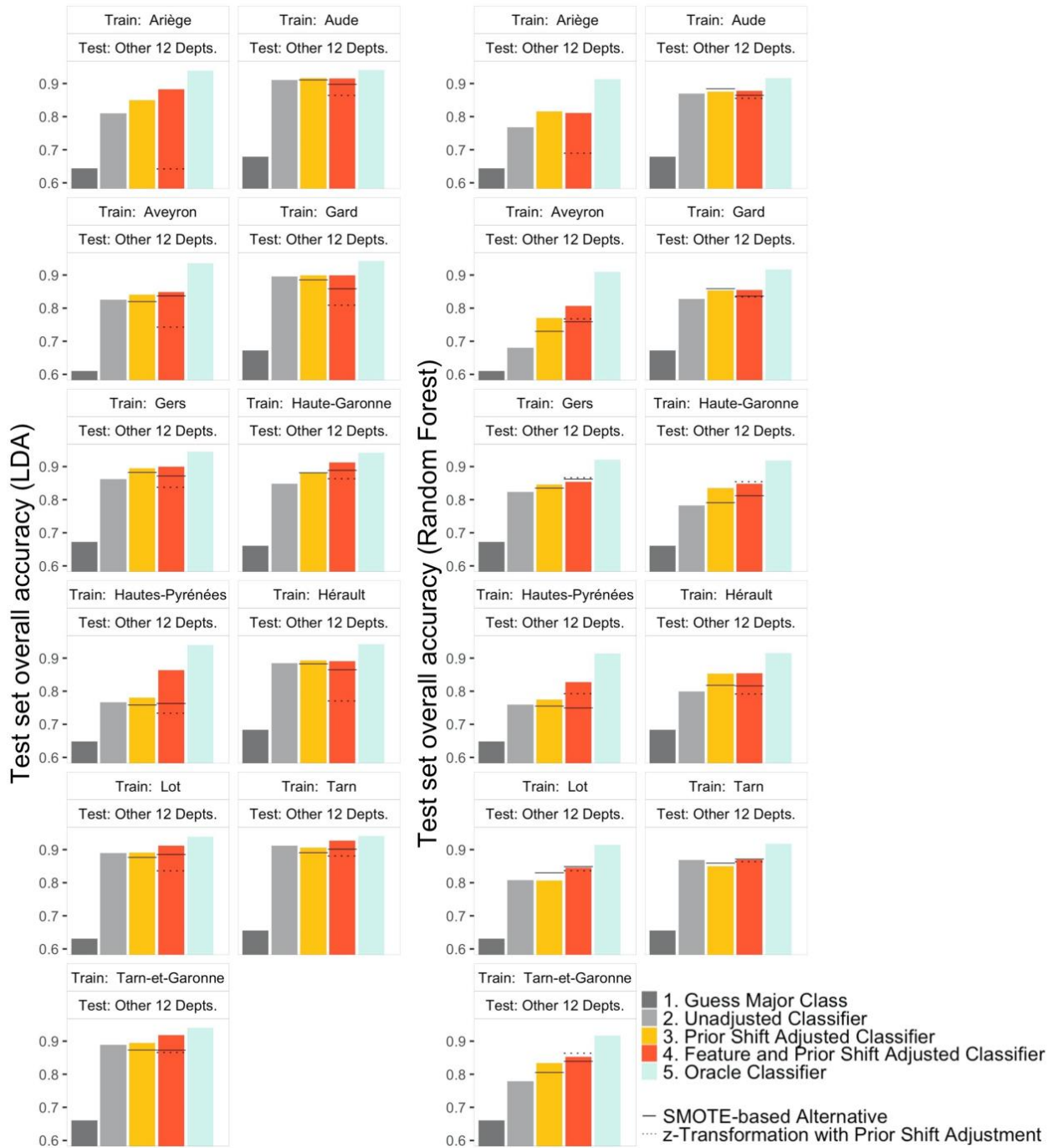

Figure 6: Barplots for the overall accuracy of the LDA (left) and Random Forest (right) classifiers when using each of the three main transfer learning approaches described (UAT, PSA, and FPSA). The barplots also show the overall accuracy when using the baseline classifier of guessing the major class. The light blue bars give the accuracy of the Oracle classifier, which is unusable in practice but gives us an estimated upper bound on the accuracy of transfer learning methods when the given base classifier is used. The barplot titles give the name of the designated training department in Occitanie, France, and the overall accuracies are calculated by applying the classifier to the 12 remaining test departments in Occitanie, France. On the barplots, we append thin dotted lines and solid lines that show the overall accuracy for alternative prior shift and feature shift adjustment methods. The solid line on the third





bar of each barplot represents the overall accuracy of SMOTE-PSA, an alternative to the PSA approach of Section 2.1. On the fourth bar, the dotted line shows the accuracy of zT-FPSA and the solid line shows the accuracy of zT-SMOTE-PSA, which are both alternative feature and prior shift adjustment approaches to our FPSA method presented in Section 2.2. Note that SMOTE with parameter k=5 could not be implemented when Ariège was the training set because Ariège only had 3 instances of wine grape.

Because overall accuracy does not give a complete picture of how well a classifier is working and can be largely driven by success of the classifier in classifying the most common class, we also use the confusion matrices from our experiments to look at other accuracy metrics. One metric we look at is the arithmetic mean of the class-wise $F_1$-scores. The metric is described in Appendix E, and Figure A3 (left panel) shows the arithmetic mean of the class-wise $F_1$-scores in each of our 11 different experiments when LDA is the base classifier. In each of the 11 experiments, we found that using the FPSA approach led to an increase in the macro-averaged $F_1$-score ranging from 0.009 to 0.159 (mean=0.063) when compared to the UAT approach while the FPSA approach led to an increase in macro-averaged $F_1$-score ranging from -0.001 to 0.114 (mean=0.026) when compared to the PSA approach.

We also look at the producer's and user's accuracies for the classifiers in our experiment. In particular, for each of the three main transfer learning approaches described (UAT, PSA, and FPSA) and for each of the six crop type classes in Occitanie, we look at the producer's accuracy and the user's accuracy averaged across the 11 experiments (where recall each experiment corresponds to a different choice of training region). These producer's accuracies and user's accuracies are presented in Table 2A, and they demonstrate the advantage of using FPSA instead of UAT or PSA.





**Table 2A:** Producer's and user's accuracies for LDA in France. Each entry is the producer's or user's accuracy averaged over the 11 experiments, where one experiment involves training the classifier on only one department and transferring the trained classifier to the remaining 12 departments. The table compares the averaged producer's and user's accuracies for the Unadjusted Transfer (UAT) classifier, the Prior Shift Adjusted (PSA) classifier, and the Feature and Prior Shift Adjusted (FPSA) classifier. The bold numbers indicate which of the three classifiers has the highest producer's or user's accuracy for that particular crop type class. We can see from the table that FPSA tends to have the highest class-wise producer's accuracies and the highest class-wise user's accuracy.

|  | Producer's Accuracies | | | User's Accuracies | | |
|---|---|---|---|---|---|---|
|  | UAT | PSA | FPSA | UAT | PSA | FPSA |
| Corn | 0.87 | 0.87 | **0.87** | 0.85 | 0.89 | **0.90** |
| Meadow-Fallow-Pastoral | 0.93 | 0.92 | **0.95** | 0.95 | **0.96** | 0.95 |
| Sunflower | 0.85 | **0.86** | 0.85 | 0.69 | 0.76 | **0.84** |
| Wine Grapes | 0.71 | 0.83 | **0.88** | 0.80 | 0.79 | **0.83** |
| Winter Barley | 0.62 | 0.65 | **0.65** | 0.46 | 0.50 | **0.52** |
| Winter Wheat | 0.71 | **0.71** | 0.70 | 0.78 | 0.82 | **0.86** |

**Table 2B:** Producer's and user's accuracies for Random Forest in France. This table was constructed in the same way as Table 2A, with the one difference being that the producer's and user's accuracies shown are for the case where Random Forest is the base classifier rather than LDA.

|  | Producer's Accuracies | | | User's Accuracies | | |
|---|---|---|---|---|---|---|
|  | UAT | PSA | FPSA | UAT | PSA | FPSA |
| Corn | 0.59 | 0.67 | **0.69** | 0.77 | **0.78** | 0.77 |
| Meadow-Fallow-Pastoral | **0.97** | 0.88 | 0.91 | 0.83 | **0.92** | 0.91 |
| Sunflower | 0.68 | **0.75** | 0.73 | 0.65 | 0.75 | **0.77** |
| Wine Grapes | 0.36 | **0.88** | 0.86 | 0.60 | 0.71 | **0.73** |
| Winter Barley | **0.17** | 0.17 | 0.17 | 0.50 | 0.57 | **0.61** |
| Winter Wheat | 0.69 | 0.69 | **0.70** | 0.76 | 0.74 | **0.78** |

### 4.2.2) With Random Forest

We also repeated the 11 experiments for the Occitanie dataset following the testing procedure described in Section 3.6, but instead of training LDA classifiers on our designated training region, we trained Random Forest classifiers. The overall accuracies are shown in Figure 6 (right





panel) and the arithmetic means of the class-wise $F_1$-scores are shown in Figure A3 (right panel).

In our 11 experiments, we found that using the FPSA approach led to an increase in overall

accuracy ranging from 0.001 to 0.127 (mean=0.049) when compared to the UAT approach.

Further, in the 11 experiments, the FPSA classifier led to an increase in overall accuracy ranging

from -0.005 to 0.053 (mean=0.017) when compared to the PSA classifier. In each of the 11

experiments, we also found that using the FPSA approach led to an increase in the macro-

averaged $F_1$-score ranging from -0.009 to 0.254 (mean=0.128) when compared to the UAT

approach while the FPSA approach led to an increase in macro-averaged $F_1$-score ranging from -

0.004 to 0.051 (mean=0.015) when compared to the PSA approach. The class-wise producer's

and user's accuracies averaged across the 11 experiments are presented in Table 2B and

demonstrate the advantage of using FPSA instead of UAT or PSA.

## 4.3) Comparison of results with other feature and prior shift adjustment approaches

We first compare the prior shift adjustment (PSA) method presented in Section 2.1 to a SMOTE-

based prior shift adjustment (SMOTE-PSA) method, which was first introduced in (Waldner *et

al.*, 2017) for binary crop classification tasks. We only describe numerical summaries of the

comparison in the setting where LDA is the base classifier and where our accuracy metric is

overall accuracy, although the comparison for Random Forest can be seen in Figure 6 and a

comparison for macro $F_1$-scores can be seen in Figure A3. From comparing the yellow bar to the

solid black line in the third column of each barplot in Figure 6 (top panel), it can be seen that the

method presented in Section 2.1 performs slightly better than the SMOTE-PSA method on our

Occitanie dataset. We note that the SMOTE-PSA method could not be implemented for all





departments with the default setting k=5, because Ariège only had 3 instances of wine grape. In the 10 remaining experiments in Occitanie, using the PSA method presented in Section 2.1 instead of the SMOTE-PSA method led to improvements in overall accuracy ranging from 0.000 to 0.022 (mean=0.014). This is understandable given that the PSA method in Section 2.1 does not rely on artificially sampling new points of under-represented crop types on the training set, as such oversampling can be unstable if there are very few instances of a particular crop type on the training set. It is also worth noting that the PSA method is computationally more efficient and easier to implement than SMOTE-PSA in multiclass classification tasks.

In Figure 6, we also compare the FPSA method presented in Section 2.2 to the two z-transformation based feature shift adjustment methods described in Section 3.8.2. Again, we only describe the comparison in the setting where LDA is the base classifier and where our accuracy metric is overall accuracy. As expected, naïvely applying z-transformations to each region without accounting for the differences in training and test set label distributions can be problematic, even when we apply a Prior Shift Adjusted classifier to the z-transformed data. In particular, we see that for the 11 experiments in Occitanie, using our FPSA approach instead of the zT-FPSA approach led to improvements in overall accuracy ranging from 0.046 to 0.241 (mean=0.093). Combining SMOTE with a z-transformation performs better, but it does not consistently perform better than the unadjusted classifier, and our FPSA approach still performs much better than this approach. In particular, we observed that for the ten experiments in Occitanie (excluding Ariège as a training region), using our FPSA method instead of zT-SMOTE-FPSA led to improvements in overall accuracy ranging from 0.012 to 0.101 (mean=0.035). Meanwhile, our FPSA method consistently leads to higher overall accuracy than





the unadjusted classifier in the examples we looked at for LDA as mentioned in Section 4.2.1.

## 4.4) Comparison of results with the Oracle Classifier

The gap between the overall accuracy of the unusable Oracle Classifier and our FPSA method ranged from 0.013 to 0.086 (mean=0.043) for LDA and from 0.038 to 0.103 (mean=0.070) for Random Forest for our 11 experiments in Occitanie. These gaps can be visualized by comparing the light blue bars with the orange bars in Figure 6.

## 4.5) Results in Western Province, Kenya

While our results in France can be helpful for checking the efficacy of our methodology on a large dataset, we now turn our attention to our results from Kenya, a setting with scarce field-level crop labels constrained to small geographical regions where such methodology is more likely to be used. We first visualize an example of feature shift in Kenya, where it appears that in Bungoma the GCVI time series tends to have a higher peak than that in Siaya for the crops banana, beans, groundnut and cassava (Figure 7). We also visualize the harmonic features in Kenya in two dimensions using t-SNE (Figure 8). T-SNE does not lead to a clear separation between classes which is a sign that attaining a high classification accuracy may be difficult on this dataset. However, if points are near each other in the t-SNE representation, they need not be near each other in the higher dimensional feature space, meaning it still may be possible to build a classifier with high classification accuracy for this dataset, despite t-SNE failing to separate the classes well.





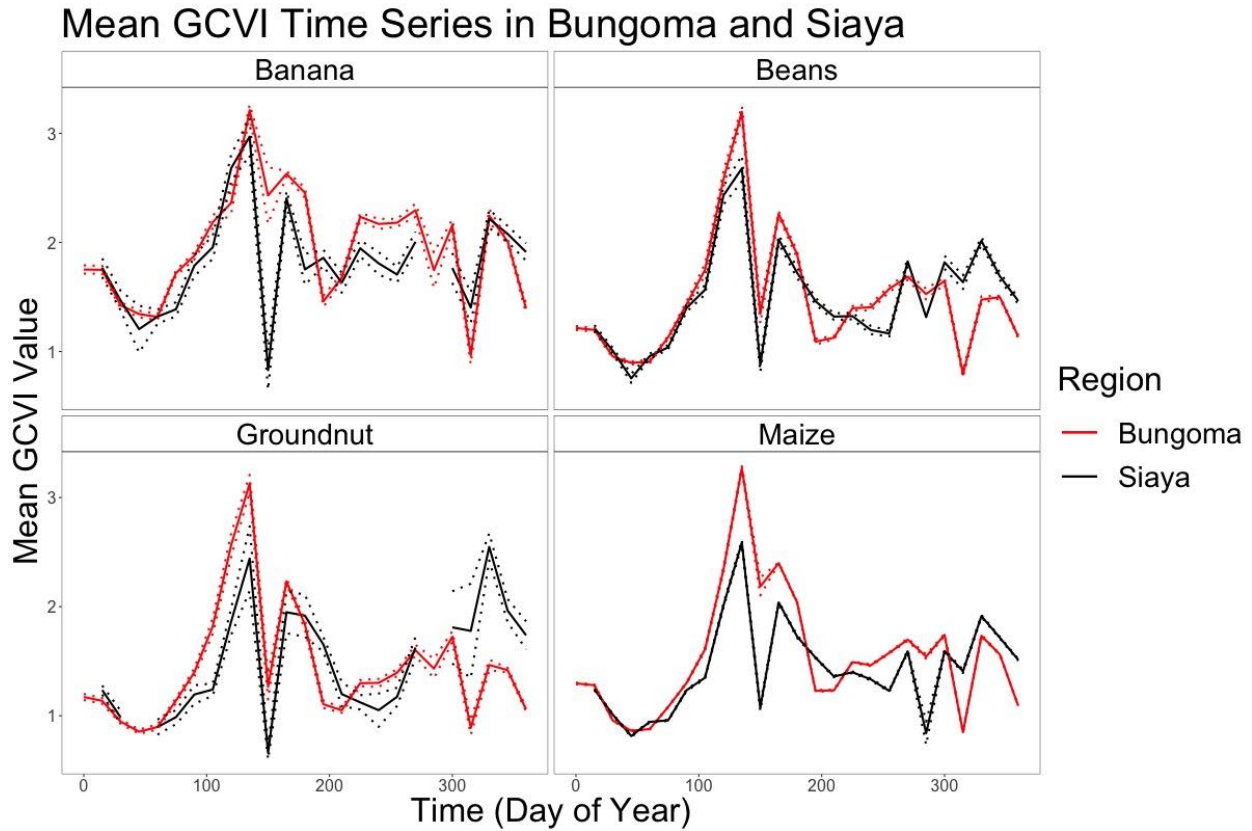

Figure 7: Plots of the GCVI time series averaged over all centroid pixels in Bungoma versus in Siaya for banana, beans, groundnut, and maize. Sentinel-2 acquisitions that were not deemed clear were excluded. We binned time into 15-day chunks before plotting averages over all clear observations within each bin. When there are no pixels that are deemed clear within a 15-day period, the plotted time series has a gap (for example, no gaps occur in the time series for more common crops such as maize, but such gaps can be seen in both the banana and groundnut time series in Siaya because there are so few pixels of these crop types in Siaya within our labeled dataset). The dotted lines give 95% confidence intervals for the mean GCVI value, although temporal correlations are ignored when defining these confidence intervals (so we should not expect 95% coverage across the time series). The abrupt decrease in the mean GCVI time series in the spring is very likely due to cloud cover that was not detected by either the QA60 measure or the Hollstein Quality Assessment measure. The plots seem to indicate that for each crop, GCVI values in Bungoma tend to attain a higher peak than those in Siaya.





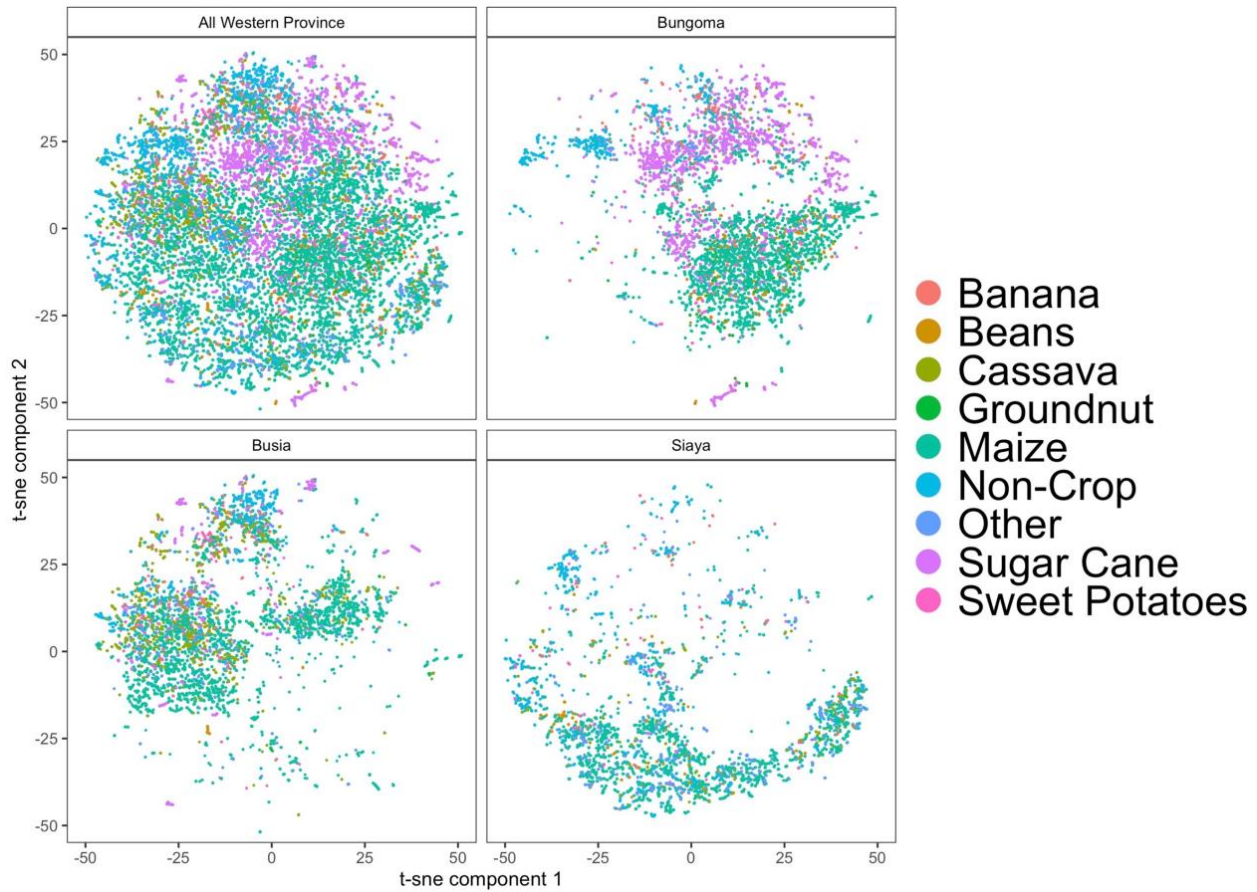

Figure 8: Visualization of all 39,553 data points (harmonic features for buffered pixels, with repeat data points removed) in Kenya using t-SNE. The different crop types are not well separated in their t-SNE representation.

The results for the three experiments in Western Province, Kenya following the testing procedure described in Section 3.6, are plotted in Figure 9. For reference we also include the accuracy of the classifier that simply guesses the most common class (maize). We found that when using LDA as a base classifier, the overall accuracy (macro-averaged $F_1$-score) improvements due to using the FPSA approach instead of the UAT approach were 0.177 (0.055), 0.025 (0.044), and 0.302 (0.115) for training on Bungoma, Busia, and Siaya respectively. The improvements in overall accuracy (macro-averaged $F_1$-score) due to using the FPSA approach instead of the PSA approach were 0.156 (0.034), 0.013 (0.029), and 0.171 (0.061) for training on these three regions.





When using Random Forest as a base classifier, the overall accuracy (macro-averaged $F_1$-score) improvements due to using the FPSA approach instead of the UAT approach were 0.021 (0.021), 0.133 (0.100), and 0.069 (0.054) for training on Bungoma, Busia, and Siaya respectively. The improvements in overall accuracy (macro-averaged $F_1$-score) due to using the FPSA approach instead of the PSA approach were 0.026 (-0.005), 0.048 (0.027), and 0.038 (-0.004) for training on these three regions.

Figure 9 also depicts the accuracies of SMOTE-PSA, zT-FPSA, and zT-SMOTE-FPSA, as well as the accuracy of the Oracle Classifier in Kenya. Figure A4 depicts the arithmetic means of the class-wise $F_1$-scores while Tables 3A and 3B present the class-wise producer's and user's accuracies for UAT, PSA, and FPSA classifiers.

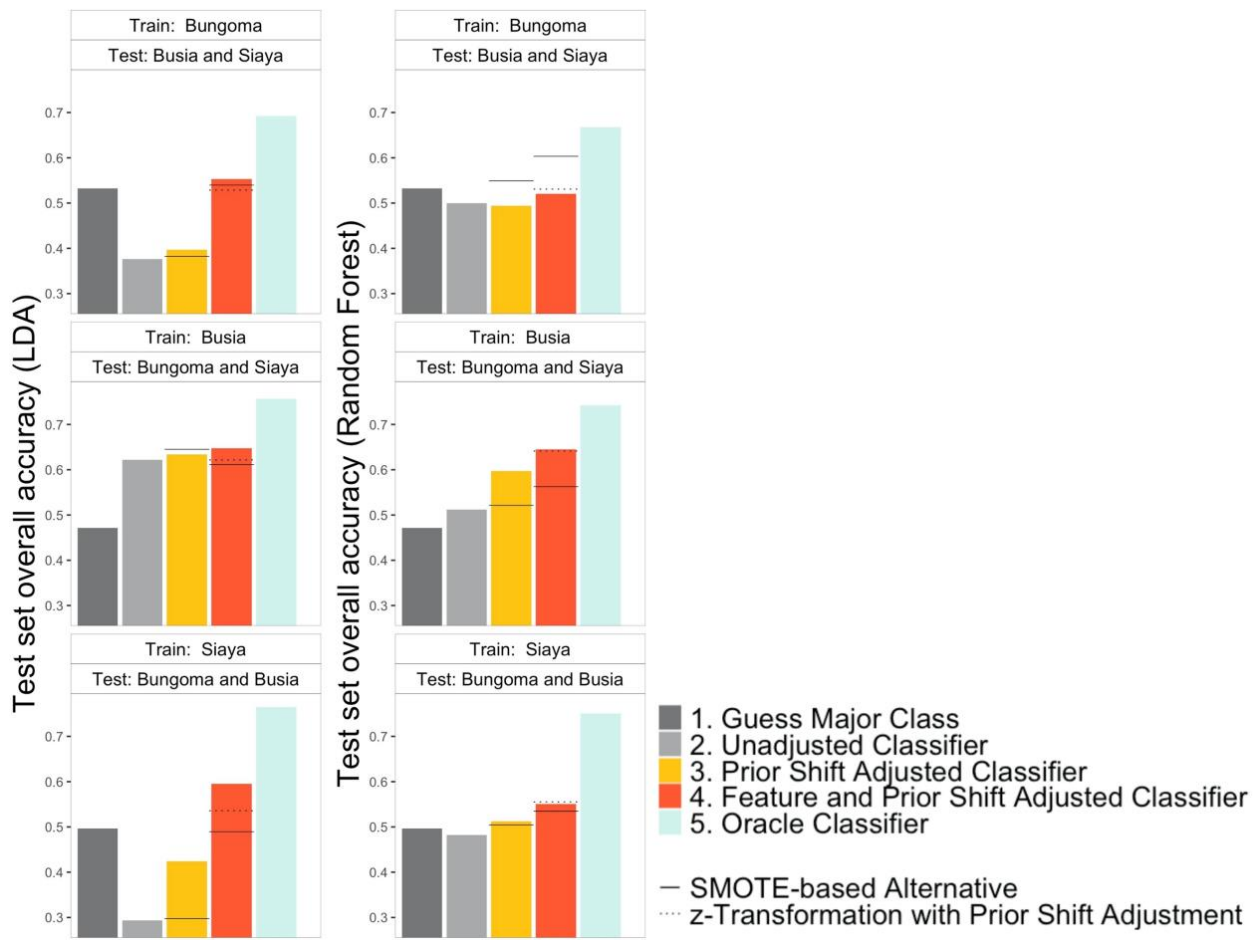





Figure 9: Barplots for the overall accuracy of the LDA (left) and Random Forest (right) classifiers when using each of the three main transfer learning approaches described (UAT, PSA, and FPSA). The barplots also show the overall accuracy when using the baseline classifier of guessing the major class. The light blue bars give the accuracy of the Oracle classifier. The barplot titles give the name of the designated training region in Western Province, Kenya, and the overall accuracies are calculated by checking their accuracy on the 2 remaining test regions. The thin dotted lines and solid lines show the overall accuracy for alternative prior shift and feature shift adjustment methods. The solid line on the third bar of each barplot represents the overall accuracy of SMOTE-PSA. On the fourth bar, the dotted line shows the accuracy of zT-FPSA and the solid line shows the accuracy of zT-SMOTE-PSA.  Note that when Bungoma was the training set, we set the SMOTE parameter to be k=4 rather than k=5, because Bungoma only had 5 labeled pixels of cassava.

**Table 3A:** Producer's and user's accuracies for LDA in Kenya. Each entry is the producer's or user's accuracy for a particular crop type class when the classifier is trained on one region and validated on the remaining two regions. The table compares the producer's and user's accuracies for the Unadjusted Transfer (UAT) classifier, the Prior Shift Adjusted (PSA) classifier, and the Feature and Prior Shift Adjusted (FPSA) classifier. The bold numbers indicate which of the three classifiers has the highest producer's or user's accuracy for that particular crop type class and training region.

| | Classifier trained in Bungoma Accuracy calculated by testing on Busia and Siaya | | | | | | Classifier trained in Busia Accuracy calculated by testing on Bungoma and Siaya | | | | | | Classifier trained in Siaya Accuracy calculated by testing on Bungoma and Busia | | | | | |
| | Producer's Accuracies | | | User's Accuracies | | | Producer's Accuracies | | | User's Accuracies | | | Producer's Accuracies | | | User's Accuracies | | |
| | UAT | PSA | FPSA | UAT | PSA | FPSA | UAT | PSA | FPSA | UAT | PSA | FPSA | UAT | PSA | FPSA | UAT | PSA | FPSA |
|---|---|---|---|---|---|---|---|---|---|---|---|---|---|---|---|---|---|---|
| Banana | **0.19** | 0.09 | 0.17 | 0.07 | **0.07** | 0.06 | 0.03 | 0.03 | **0.04** | **0.11** | 0.09 | 0.06 | 0.14 | **0.22** | 0.11 | 0.09 | 0.07 | **0.10** |
| Beans | **0.23** | 0.11 | 0.06 | 0.05 | 0.06 | **0.08** | 0.00 | 0.00 | **0.05** | 0.00 | 0.00 | 0.00 | 0.08 | 0.08 | 0.04 | 0.01 | 0.02 | **0.05** |
| Cassava | 0.15 | **0.37** | 0.23 | **0.33** | 0.21 | 0.28 | **0.43** | 0.32 | 0.29 | 0.27 | 0.47 | **0.52** | 0.16 | 0.28 | **0.40** | **0.65** | 0.58 | 0.55 |
| Groundnut | 0.00 | 0.00 | 0.00 | 0.00 | 0.00 | 0.00 | 0.00 | **0.01** | 0.00 | 0.00 | **0.00** | 0.00 | 0.00 | 0.00 | 0.00 | 0.00 | 0.00 | 0.00 |
| Maize | 0.43 | 0.42 | **0.77** | 0.72 | **0.76** | 0.70 | **0.93** | 0.89 | 0.83 | 0.61 | 0.64 | **0.70** | 0.35 | 0.55 | **0.80** | **0.71** | 0.70 | 0.69 |
| Non-Crop | **0.48** | 0.47 | 0.48 | 0.54 | 0.56 | **0.61** | 0.39 | 0.40 | **0.54** | **0.82** | 0.82 | 0.64 | **0.76** | 0.68 | 0.48 | 0.22 | 0.26 | **0.45** |
| Other | 0.10 | **0.26** | 0.14 | 0.13 | **0.16** | 0.14 | 0.03 | 0.04 | **0.09** | 0.09 | 0.10 | **0.13** | **0.09** | 0.04 | 0.06 | 0.02 | 0.02 | **0.07** |
| Sugar Cane | **0.54** | 0.35 | 0.42 | 0.33 | **0.63** | 0.62 | 0.51 | 0.65 | **0.73** | **0.86** | 0.81 | 0.79 | 0.13 | 0.25 | **0.51** | **0.91** | 0.86 | 0.70 |
| Sweet Potatoes | **0.58** | 0.44 | 0.27 | 0.06 | 0.06 | **0.20** | 0.15 | 0.15 | **0.25** | **0.71** | 0.68 | 0.37 | 0.33 | **0.34** | 0.32 | 0.13 | 0.12 | **0.14** |

**Table 3B:** Producer's and user's accuracies for Random Forest in Kenya. This table was constructed in the same way as Table 3A, with the one difference being that the producer's and user's accuracies shown are for the case where Random Forest is the base classifier rather than LDA.

| | Classifier Trained in Bungoma Accuracy calculated by testing on Busia and Siaya | | | | | | Classifier Trained in Busia Accuracy calculated by testing on Bungoma and Siaya | | | | | | Classifier Trained in Siaya Accuracy calculated by testing on Bungoma and Busia | | | | | |
| | Producer's Accuracies | | | User's Accuracies | | | Producer's Accuracies | | | User's Accuracies | | | Producer's Accuracies | | | User's Accuracies | | |
| | UAT | PSA | FPSA | UAT | PSA | FPSA | UAT | PSA | FPSA | UAT | PSA | FPSA | UAT | PSA | FPSA | UAT | PSA | FPSA |
|---|---|---|---|---|---|---|---|---|---|---|---|---|---|---|---|---|---|---|
| Banana | **0.04** | 0.00 | 0.00 | **0.05** | 0.00 | 0.00 | 0.00 | **0.00** | **0.00** | 0.00 | 0.06 | 0.04 | 0.00 | **0.01** | 0.00 | **1.00** | 0.15 | 0.33 |
| Beans | 0.00 | 0.00 | 0.00 | 0.00 | 0.00 | 0.00 | 0.00 | 0.00 | 0.00 | 0.00 | 0.00 | 0.00 | 0.00 | 0.00 | 0.00 | 0.00 | 0.00 | 0.00 |
| Cassava | 0.00 | **0.33** | 0.22 | 0.00 | **0.18** | 0.14 | **0.17** | 0.00 | 0.00 | 0.05 | 0.00 | **0.33** | 0.00 | **0.43** | 0.24 | 0.00 | 0.22 | **0.36** |
| Groundnut | 0.00 | 0.00 | 0.00 | 0.00 | 0.00 | 0.00 | 0.00 | 0.00 | 0.00 | 0.00 | 0.00 | 0.00 | 0.00 | 0.00 | 0.00 | 0.00 | 0.00 | 0.00 |
| Maize | 0.72 | 0.69 | **0.76** | 0.69 | **0.69** | 0.67 | **0.98** | 0.89 | 0.85 | 0.52 | 0.58 | **0.66** | **0.81** | 0.61 | 0.77 | 0.63 | **0.72** | 0.65 |
| Non-Crop | 0.47 | 0.44 | **0.48** | 0.50 | 0.48 | **0.52** | 0.20 | 0.41 | **0.58** | **0.73** | 0.73 | 0.54 | **0.92** | 0.13 | 0.10 | 0.24 | 0.57 | **0.66** |
| Other | 0.00 | **0.13** | 0.10 | 0.09 | 0.11 | **0.14** | 0.00 | **0.06** | 0.00 | 0.09 | **0.18** | 0.17 | **0.03** | 0.00 | 0.00 | **0.03** | 0.00 | 0.00 |
| Sugar Cane | **0.79** | 0.13 | 0.11 | 0.14 | 0.57 | **0.58** | 0.08 | 0.51 | **0.71** | **0.96** | 0.66 | 0.74 | 0.00 | **0.73** | 0.61 | **1.00** | 0.39 | 0.40 |
| Sweet Potatoes | **0.02** | 0.01 | 0.00 | **0.32** | 0.14 | 0.00 | 0.01 | 0.00 | **0.01** | **0.57** | 0.00 | 0.09 | **0.05** | 0.03 | 0.02 | **0.35** | 0.17 | 0.28 |

The results of Tables 3A and 3B show that in Kenya, neither UAT, PSA, nor FPSA consistently performs the best across every crop type; however, FPSA still performs the best when considering the overall accuracy and when considering the macro-averaged $F_1$-score. Tables 3A





and 3B also highlight that PSA and FPSA account for the crop type distributions in the target

regions. For example, consider the scenario when the classifiers are trained in Bungoma (left

columns in Tables 3A and 3B). Bungoma has much larger percentage of sugar cane and a much

smaller percentage of cassava than the other two regions (see Figure 3). As a result, we observe

that when training in Bungoma, the UAT classifier guesses sugar cane more frequently than the

PSA and FPSA classifiers guess sugar cane, and conversely, the UAT classifier guesses cassava

less frequently than the PSA and FPSA classifiers guess cassava.

## **5) Discussion**

In Section 5.1 we provide an overview of the results. In Sections 5.2 we discuss the suitability of

the FPSA method to the setting where LDA is the base classifier, while in Section 5.3 we discuss

its suitability when Random Forest or another non-LDA classifier is the base classifier,

suggesting directions for future research to improve upon the FPSA method. In Section 5.4 we

discuss the benefits of picking a region with diverse crop types to train on if the option is

available.

## **5.1) Overview of Results**

Our results indicate a clear benefit of using FPSA over the naïve unadjusted classifier (UAT) in

both France and in Kenya. In France, the benefit of FPSA over UAT was clearly demonstrated

when looking at either overall accuracy, macro-averaged class-wise $F_1$-scores, producer's

accuracies or user's accuracies. The benefits of using FPSA rather than UAT varied based on the

choice of training department. For example, when Tarn was the training region, the FPSA

classifier did not demonstrate a substantial benefit over the UAT classifier. Meanwhile, when the

training region was Hautes Pyrénées, FPSA led to a substantial improvement over UAT. This

observation can be partially explained by two reasons. First, the distribution of the crop types in





Tarn is very similar to that of Occitanie as a whole, whereas the distribution of the crop types in Hautes Pyrénées is not so similar to that of Occitanie, implying that prior shift adjustments would be much more substantial and beneficial for classifiers trained in the latter region. Second, Tarn is geographically in the center of Occitanie, whereas Hautes Pyrénées is in the southwest corner of the region, implying that feature shift adjustments for classifiers trained in the latter department would be much more substantial. On a similar note, we observed that the overall accuracies varied (when using FSPA) based on the training department. In particular, when the classifier was trained in regions with one main class and few instances of the remaining classes, such as Aveyron, the overall accuracy tended to be lower. This phenomenon is investigated more thoroughly and explained in Section 5.4.

The overall accuracy improvements of using FPSA rather than UAT tended to be particularly large in Kenya, and much larger than those in France. This is largely driven by the fact that in France the baseline UAT classifier already performed quite well, so there was less room for feature and prior shift methods to confer improvements in terms of overall accuracy. It should be noted that large obstacles still exist for creating highly accurate crop type maps in Kenya with the available data. Even with LDA as a base classifier our FPSA method led to overall accuracies in the 0.55 to 0.65 range in Kenya, and when Bungoma was the training region, the FPSA classifier was not much more accurate than the approach of always guessing the major class.

The lower accuracies in Kenya are due to a few factors that make our classification task in Kenya more difficult. First, intercropping is common practice in Kenya making crop classification in Kenya more difficult. Second, we use Sentinel-2 Level-1C cloud masks products, which have been observed to preform worse in non-Mediterranean climates (Coluzzi *et al*., 2018). Third, in





Western Kenya there is high climatic variability even within small geographic regions such as the Western Province. Fourth, we included "other" crops and non-cropped land in our classification task in Kenya. The fact that in Kenya we often observed guessing the major class to lead to higher overall accuracy than UAT or even PSA is a manifestation of the difficulty of the Sentinel-2-based classification task in Kenya. FPSA, however, still generally led to better overall accuracy than guessing the major class. It is also interesting to note that when using UAT in Kenya, LDA tended to preform worse than Random Forest, whereas LDA with FPSA tended to perform better than Random Forest with UAT or FPSA, suggesting that the LDA base classifier should not be ruled out simply because it has low accuracy when using UAT.

In addition to observing that FPSA tended to confer larger benefits over UAT when LDA, rather than Random Forest, was the base classifier, there are some other notable differences between our results for these two base classifiers. We also found that the advantage of using FPSA over other transfer approaches holds consistently when LDA is the base classifier. It also typically, but less consistently, holds when Random Forest is the base classifier. Further, the gap between the FPSA and the Oracle accuracies tended to be much larger for Random Forest. We provide a possible explanation for these phenomena in Sections 5.2 and 5.3. It should also be noted that the large gaps between the Oracle accuracies and FPSA accuracies suggest that it might be possible to develop transfer learning method approaches that preform substantially better than FPSA.

## 5.2) Prior and Feature Shift Adjustment for Linear Discriminant Analysis

Based on the results, the FPSA methodology presented in Section 2 performs well relative to the baseline and alternative approaches that we looked at for LDA. It always performed better than the six other approaches, with the exception that in 3 of the 11 experiments in Occitanie the





overall accuracy of our FPSA method was between 0.000 and 0.002 lower than that for the PSA method.

LDA is a base classifier that is particularly well suited for our methodology for two reasons. First, the posterior probabilities returned by the classifier have a probabilistic interpretation: they are the probabilities of being in each class conditional on the features for a fitted multivariate Gaussian mixture model with common covariance. Therefore, if the data approximately follows a multivariate Gaussian mixture model with common covariance, and if the training set is large enough, LDA will return very accurate posterior probabilities. Accurate posterior probabilities are beneficial to the PSA classifier we presented in Section 2.1 because the PSA classifier uses reweighted versions of these posterior probabilities to fit labels on the test set. Second, the feature means for each class are critical components of an LDA classifier. Recall that when applying an LDA classifier to a test point, for each class we must calculate the density at that point of a multivariate Gaussian with mean vector equal to the mean feature vector for that class. Therefore, if there is a shift in the feature means for each class between regions, it can be quite detrimental to the LDA classifier. Our methodology in Section 2.2 removes shifts in the feature means for each class between regions under the assumption that the feature mean shifts for each crop type is the same.

### 5.3) Prior and Feature Shift Adjustment for non-LDA classifiers





The prior shift adjustment method presented in Section 2.1 should work well for any choice of classifier, while the feature shift adjustment method developed in Section 2.2 can be helpful for certain classifiers. Our results for Random Forest suggest that FPSA is helpful compared to UAT, but they also suggest that FPSA can be improved upon and that it is worth researching robust alternatives to FPSA for non-LDA classifiers. For example, in Western Kenya, when the training set is from Bungoma, Kenya, the FPSA method dominates the SMOTE-based methods for LDA but not for Random Forest. In addition, the gap between the FPSA accuracies and the Oracle accuracy tended to be larger for Random Forest than for LDA.

The "posterior probabilities" of being in each class given the features returned by a Random Forest are based on the proportion of trees that vote for each class and therefore have no probabilistic motivation. In comparison, for LDA, the posterior probabilities returned by the classifier have a probabilistic interpretation as described in Sections 3.5 and 5.1. Still, even if the posterior probabilities for a Random Forest are more poorly calibrated than those for LDA, the method in Section 2.1 will reweight these somewhat inaccurate probabilities by the proper factor, which often leads to classification improvements for Random Forest.

If the true feature transformation between regions is simply a translation, then our method presented in Section 2.2 directly removes the issues associated with feature shift regardless of the base classifier being used. In the more realistic scenario, when the feature transformation between regions is not exactly a translation, our FSA and FPSA methods can help for any type of classifier, although it is understandable that such methods might confer less benefit when using a non-LDA base classifier. Recall that our method presented in Section 2.2 adjusts the features on





the testing set so that the class-wise feature means for the training and test set are consistent. Because LDA assigns a label to a training point by picking the class with the nearest class-wise mean feature vector (in terms of Mahalanobis distance), having consistent class-wise feature means on the training set and test set is critical for LDA classification. For most non-LDA classifiers, including Random Forests, the class-wise feature means are not directly used in assigning a fitted class label, so having consistent class-wise feature means between the training and test set is helpful, but not necessarily the most importing thing to ensure.

As deep learning methods are becoming more widely used for crop classification, future work should explore how effective our proposed PSA and FPSA methods are when the base classifier is a Neural Network. Both PSA and FPSA can easily be applied post-hoc when a Neural Network is trained for crop classification. When the feature space involves raw multispectral time series inputs rather than the harmonic features considered in this paper, the mean feature shift would be computed using the method described in Section 2.2 with each date and spectral band pair being a unique entry in the feature vector. In addition, two practical considerations for using PSA or FPSA are worth mentioning. First, one should ascertain that the original Neural Network classifier is trained with a cross-entropy loss so that the posterior probabilities returned by the classifier have the potential to be well calibrated. Second, when estimating the mean feature shift of a raw multispectral time series, cloudy days should be removed. Similar to our observations for Random Forest, we anticipate that FPSA will lead to accuracy improvements for Neural Networks transferred across regions because the method will still reweight the estimated posterior probabilities to favor guessing prevalent classes in the target region and will remove mean feature shifts; however, its efficacy should be checked empirically in future work because





the posterior probabilities in Neural Networks can be poorly calibrated and because Neural

Networks apply many nonlinear transformations to the features.

In summary, our FSA and FPSA methods correct for feature mean shifts, which are the most

important thing to correct for in a classifier like LDA. Feature mean shift is also the only thing

that needs to be corrected if the systemic feature transformation between regions is simply a

translation. However, for different classifiers such as Random Forests or Neural Networks, in

order to have a reliable and highly beneficial feature transformation adjustment, we may need to

correct for feature transformations other than mean shift. Therefore, a future direction would be

to anticipate the types of systemic feature transformations that occur between regions. For

example, in our setting of crop classification using harmonic regression features, we might

expect the growing season to shift by a few days or weeks between different regions leading to a

phase shift in the time series. A phase shift would correspond to a rotation of each pair of

harmonic features corresponding to the same frequency and spectral band by an angle

proportional to the product of the phase shift and the frequency (see Appendix D for a

mathematical justification of this claim). Alternatively, if the amplitude of the time series

changes between regions, this would lead to a translation and possibly a scaling of the harmonic

features. It is less clear how to characterize the effect of shrunken or expanded growing seasons

on the harmonic features, although these shifts are also likely to occur between regions (see

Appendix D for a more detailed discussion and justification of these points). Knowing the type of

systemic feature transformations that occur between regions can motivate improved feature shift

methodologies that go beyond correcting for feature mean shift.





## 5.4) Selecting the optimal training region

A relevant question for crop type mapping is where to focus collection of field-level data among several possible regions when resources are limited. All other factors being equal, one would want to choose the region that would lead to highest overall accuracy in nearby regions. Aggregate-level data can be useful in gauging which region, if chosen as the training region, would lead to higher overall accuracy. In particular, we hypothesize that a high crop-diversity region would be ideal to label because a) it is best to train a classifier on data where there are many examples of each crop type that appears frequently in the target regions and b) after prior shift adjustments, it is much more difficult to correctly guess the labels in a high diversity region (so including a high diversity region in the training set will remove that region from the list of unlabeled regions). To test this, in Figure 10 we plot the Shannon Entropy (a crop diversity measure used in Wang *et al.*, 2019) of the training set versus the test set accuracy in France to see if this pattern holds. Shannon Entropy is given by $-\sum_k \pi_k \log(\pi_k)$ where $\pi_k$ is the proportion of crop type $k$ on the training set. Shannon Entropy is a reasonable diversity measure, because it obtains its minimum possible value of 0 whenever at least one crop type is not present, and it is maximized when each crop type has the same prevalence. Indeed, we observe that diversity of the training set was correlated with overall accuracy after adjustments in our 11 experiments in France. Further, we looked at accuracies on different regions making up the test set and found that the accuracy tended to be higher on less diverse regions within the test set. This is consistent with our hypothesis that for two reasons (sufficient training labels for each class and less diverse testing regions) it is beneficial to train on a more diverse region.





It is important to note that the benefits conferred by training on a diverse region can easily fail to hold if we ultimately neglect to use a classifier with prior shift adjustments. For example, imagine you have one diverse department with a small proportion of corn and winter barley and that the remaining nearby departments are nearly 50% corn and nearly 50% winter barley. If ultimately one neglects to use a classifier with prior shift adjustments, then it would be better to just train on one of the more representative departments (with about 50% corn and 50% winter barely) than to train on the diverse department. On the other hand, if we actually use a classifier with prior shift adjustments (such as FPSA), then it would be better to collect labels for training on the diverse department as the prior shift adjustments would allow it to successfully transfer to regions with much more corn and winter barley, while having the actual labels on the diverse department would mitigate the need to confront the more difficult task of guessing crop type labels on a diverse region with a classifier trained in a nearby region. In summary, assuming all other cost factors are equal, it would be beneficial to collect labels for training a crop classifier on a region with high crop type diversity if prior shift adjustment is used.





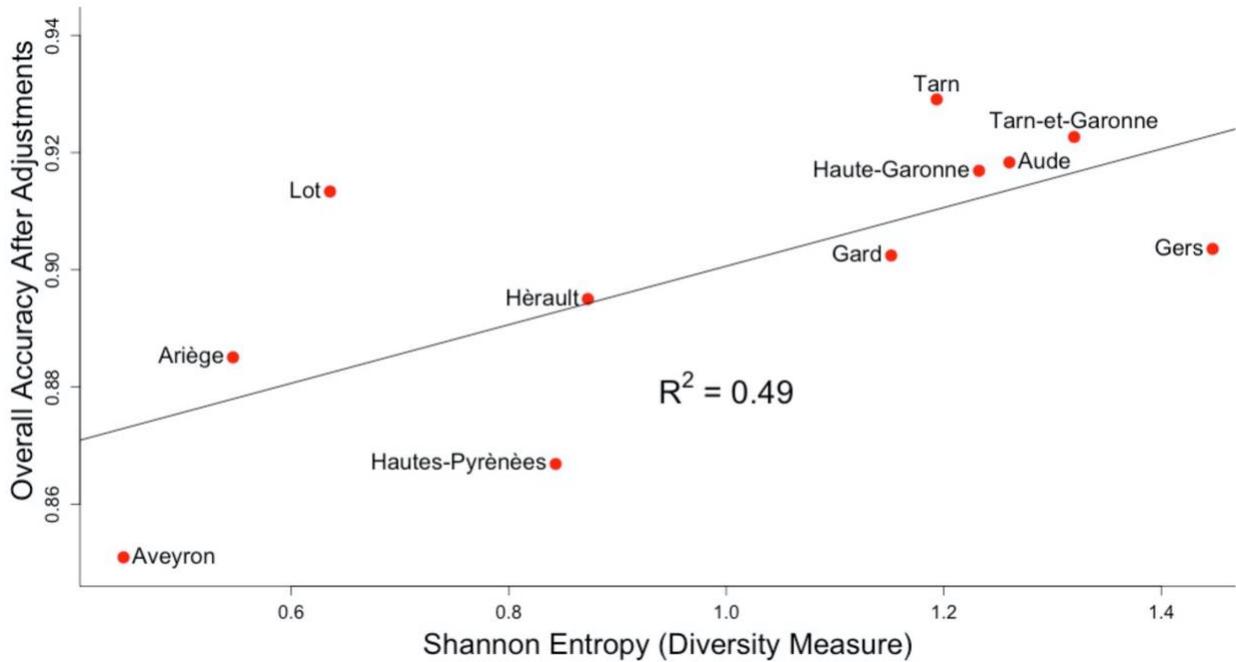

Figure 10: Diversity of the training region versus overall accuracy after applying our FPSA methodology to an LDA classifier. The name above each point designates the training region for that experiment and the x-axis value is the Shannon Entropy of the distribution of class labels in that training region.

## 6) Conclusion

In this study, we proposed and assessed an approach for transferring classifiers to regions with shifted features and different crop type compositions than the original training region. In particular, to correct for differing crop type compositions in the training region and the testing region we present a method from the machine learning literature to reweight the classifier's posterior probabilities for each class by the ratio of that class's prevalence on the test set over its prevalence on the training set. To correct for feature shifts between regions, we propose a method to estimate and then adjust for the component of the shift in the mean feature vector that is not attributable to changing class composition between regions. On two example datasets, we found that this methodology led to clear and often substantial improvements in overall accuracy when using LDA as our base classifier. We also demonstrated that the methodology can lead to





improvements in overall accuracy for other classifiers, focusing on Random Forests as an example. Another advantage of the PSA and FPSA methods is that when creating crop type maps, the methods can be applied to an already-trained classifier. In particular, to create a crop type map in a new region using FPSA with a classifier already trained in a nearby region, one must collect all of the agricultural pixels in the new target region, estimate the additive regional effect as in Section 2.2, subtract the regional effect from all pixels, apply the classifier, reweight the returned posterior probabilities as described in Section 2.1, and for each pixel classify it as the crop type with the highest reweighted posterior probability.

There are a few practical considerations to consider for operational implementation of our methods. First, while the methodology demonstrates success and promise for mapping past crop seasons, its use of a full-year time series and aggregate-level government statistics render it less beneficial for near real time mapping. That being said, it may still be possible for PSA or FPSA to be successfully implemented for real time crop mapping using a different set of early growing season features and using historical data on the crop type composition in each region to estimate the distribution of crop type classes on the unlabeled data for the current year. Second, if the collected ground labels are not a random sample of pixels in a region—for example, if they are collected with roadside sampling —then one should also consider using PSA or FPSA to create satellite-based crop type maps in the *same* region if it is thought that certain crop types are underrepresented or overrepresented in roadside sampling. With roadside sampling, PSA and FPSA are suitable for creating satellite-based crop type maps in the nearby regions as long as the training set crop type probabilities used in PSA or FPSA are computed using the actual labels on the gathered training data rather than the training region's aggregate-level statistics. Third, FPSA should still confer benefits for any choice of preprocessing steps, cloud masking procedure,





features, and base classifier. Therefore, even if the classifiers presented in this manuscript could be improved upon had we used top-of-canopy corrections, a different cloud-mask, different features than harmonic features, and a different choice of base classifier than Random Forest or LDA, FPSA would still confer additional benefits on top of the improved classifier. Fourth, it is important to note that the methodology relies on aggregate crop statistics at the regional level. These aggregate-level statistics only need to be accurate in a relative sense as our approach simply requires a good estimate of the proportion of unlabeled data points that are of each crop type. As aggregate-level statistics at the regional level become more widely available, the methodology can become more broadly applicable and can lead to a reduction in the number of regions where expensive field surveys are needed for creating automated crop type maps with remotely sensed satellite data.

One future direction for this work is to explore whether the PSA and FPSA methods are effective in transferring classifiers across years rather than across geographical regions. Another future direction would be to characterize the types of systemic feature transformations that occur between regions, beyond just feature translations, and to develop methods to correct for these types of feature transformations. We suspect that correcting for the true feature transformations that occur between regions will lead to better classification performance than a correction for feature translations, especially for non-LDA classifiers, such as the increasingly popular Neural Networks.





## Appendix

The appendix can be viewed using the following link:

https://drive.google.com/file/d/1GAqa81d6BGb5G8zwC3bzFytXPI-X6hxI/view?usp=sharing

## Acknowledgements

The authors wish to thank Trevor Hastie and Paul Switzer for helpful discussions. The authors would also like to thank the associate editor and 3 anonymous reviewers for comments that helped improve the paper. DMK was supported by a Stanford Graduate Fellowship during the research and writing of this manuscript.